\documentclass[
 reprint,
 amsmath,amssymb,
 aps,nofootinbib
]{revtex4-2}

\usepackage[utf8]{inputenc}
\usepackage{graphicx}
\usepackage{dcolumn}
\usepackage{bm}
\usepackage{hyperref}
\usepackage{bm}
\usepackage{lipsum}
\usepackage{xcolor}
\usepackage{calc}
\usepackage{accents}
\usepackage{comment}
\usepackage{float}
\newcommand{\Odcdm}{\Omega_{\rm{dcdm}}^{\rm{ini}} }

\newcommand{\eps}{\varepsilon}
\newcommand\wdm{{\rm{wdm}}}
\newcommand\dr{{\rm{dr}}}
\newcommand\dcdm{{\rm{dcdm}}}

\begin{document}

\preprint{APS/123-QED}

\title{Linear cosmological constraints on 2-body decaying dark matter scenarios \\ and
the $S_8$ tension}
\author{Guillermo F. Abell\'an}
\email{Electronic address: guillermo.franco-abellan@umontpellier.fr }
\author{Riccardo Murgia}
\email{Electronic address:  riccardo.murgia@umontpellier.fr}
\author{Vivian Poulin}
\email{Electronic address: vivian.poulin@umontpellier.fr }

\affiliation{
 Laboratoire Univers \& Particules de Montpellier (LUPM), Universit\'e de Montpellier (UMR-5299) \\ Place Eugène Bataillon, F-34095 Montpellier Cedex 05, France 
}

\date{\today}

\begin{abstract}

The `$S_8$ tension' is a longstanding discrepancy between the { cosmic microwave background (CMB) and weak gravitational lensing} determination of the amplitude of matter fluctuations, parametrized as $S_8\equiv\sigma_8(\Omega_m/0.3)^{0.5}$, where $\sigma_8$ is the root mean square of matter fluctuations on a 8 $h^{-1}$Mpc scale, and $\Omega_m$ is the total matter abundance. 
It was recently shown that dark matter (DM) decaying into a massless (dark radiation) and a massive (warm DM) species, with a lifetime $\Gamma^{-1} \simeq 55~ (\varepsilon/0.007)^{1.4}$ Gyrs -- where $\varepsilon$ represent the mass-energy fraction transferred to the massless component -- can { ease} the tension. Thanks to a fast and accurate fluid approximation scheme for the warm species, we perform a comprehensive study of this 2-body decaying DM scenario, discussing in detail its dynamics and its impact on the CMB and linear matter power spectra. We then
{investigate the implications for the `$S_8$ tension' against}
a number of changes in the analysis: different $S_8$ priors, marginalization over the lensing information in~{\emph{Planck}} data, trading {\emph{Planck}} high$-\ell$ polarization data for those from the SPTpol {and ACTPol surveys}, and the inclusion of the recent results from the Xenon1T collaboration. We conclude that the preference for decaying DM, 
{apparent only when the $S_8$ value determined from weak lensing data is added to the analysis},
does not sensibly degrade the fit to any of the cosmological data-sets considered, and that the model could {potentially} explain the anomalous electron recoil excess reported by the Xenon1T collaboration. {Furthermore, we explictly show that while current CMB data alone are not sensitive enough to distinguish between standard CDM and  decaying DM, next-generation CMB observations (CMB-S4) can unambiguously detect its signature.}

\end{abstract}

\maketitle

\section{\label{sec_dcdm:Intro}Introduction}
In the last couple of decades, the so-called standard $\Lambda$ Cold Dark Matter ($\Lambda$CDM) model of cosmology has been firmly established as the most successful framework to interpret numerous independent experimental observations up to a very high degree of accuracy. Still, the nature of its dominant components -  Cold Dark Matter (CDM) and Dark Energy (DE) -- is yet to be unveiled.
The $\Lambda$CDM model provides indeed a remarkable fit to a wide variety of early universe data, such as Cosmic Microwave Background (CMB) and Big Bang Nucleosynthesis (BBN), as well as late universe observations 
such as 
Baryon Acoustic Oscillation (BAO), and luminosity distance to SuperNovae of type Ia (SNIa). 
However, as the accuracy of the measurements has increased over the past few years, a number of intriguing discrepancies between the values of some cosmological parameter as inferred within $\Lambda$CDM and their direct measurements at low redshift, has emerged.  At the heart of this study is the mild ($\sim 2 - 3 \sigma$) yet longstanding tension between the {CMB~\cite{Aghanim:2018eyx} and weak gravitational lensing~\cite{Abbott:2017wau,Hildebrandt:2018yau,Joudaki:2019pmv}} determination of the amplitude of the matter fluctuations, tipically parametrized -- {in the context of weak lensing surveys} -- as $S_8\equiv\sigma_8(\Omega_m/0.3)^{0.5}$, where $\sigma_8$ is the root mean square of matter fluctuations on a 8 $h^{-1}$Mpc scale, and $\Omega_m$ is the total matter abundance.
Another example is the so-called Hubble tension ~\cite{Verde:2019ivm, Aylor:2018drw, Wong:2019kwg, Freedman:2019jwv,Riess:2020fzl} ,~i.e.,~the significant discrepancy ($\sim 4~-~5~ \sigma $ C.L.) between the value of the current expansion rate of the universe $H_0$, directly measured using SNIa data as a cosmic distance ladder \cite{Riess:2019cxk}, and that inferred from CMB data
\cite{Aghanim:2018eyx}, assuming $\Lambda$CDM. 
Despite scrupulous efforts to minimize the systematic errors at play in the cepheid calibration of the SNIa, the statistical significance of the disagreement is steadily rising~\cite{Shanks:2018rka,Riess:2018kzi,Davis:2019wet,Yuan:2019npk,Riess:2020fzl}. 
However, it is still debated whether the Hubble constant resulting from a calibration of the SNIa on the `tip of the red giant branch' shows a similar degree of tension with respect to {\emph{Planck}} $\Lambda$CDM model \cite{Freedman:2019jwv,Yuan:2019npk,Cerny:2020inj,Soltis:2020gpl}. Alternative methods have been proposed, but are currently not at the accuracy level required to unambiguously weigh in the Hubble tension~\cite{Verde:2019ivm}.  

In the absence of convincing solutions within the standard cosmological model, and driven by the fact that the nature of the dark sector is unknown, throughout the years many alternative scenarios have been proposed to explain these discrepancies. It is {not yet clear if introducing new physics in the pre-recombination era to decrease the sound horizon at recombination could fully resolve the Hubble tension without spoiling other bounds or exacerbating the $S_8$ tension~\cite{Knox:2019rjx,Jedamzik:2020zmd,Haridasu:2020pms,DiValentino:2021izs,Vagnozzi:2021gjh}.} 
On the other hand, resolving the $S_8$ tension requires to decrease the amplitude of matter fluctuations on scales $k\sim 0.1-1 ~h$/Mpc, which can be easily achieved in a variety of models {departing from $\Lambda$CDM only at late times}, often related to new DM properties \cite{Kumar:2016zpg,Murgia:2016ccp,Archidiacono:2019wdp,DiValentino:2020vvd}, including the possibility that DM decays \cite{Enqvist:2015ara,Poulin:2016nat,vattis_late_2019,Haridasu:2020xaa,Clark:2020miy}.

In this paper, we reassess the phenomenology of a 2-body Decaying Cold Dark Matter (DCDM) scenario, where the decay products are one massive Warm DM (WDM) particle and one (massless) DR component, interacting only through gravitation with the standard model particles. We will refer to the full model as $\Lambda$DDM. 
From the point of view of particle physics model building, the stability over cosmological timescales is one of the most peculiar property of the dark matter particle, reviewed e.g. in Ref. \cite{Hambye:2010zb}. Often, an additional symmetry (typically a discrete $Z_2$ symmetry) has to be assumed to make the DM candidate stable. 
Nevertheless, DM decays at late-times are known signatures of many models in the literature such as (for instance) models with R-parity violation \cite{BEREZINSKY1991382,KIM200218},  super Weakly Interacting Massive particles (super WIMPs)  \cite{CoviEtAl1999,FengEtAl2003,FengEtAl2003b, AllahverdiEtAl2015},  sterile neutrinos \cite{Abazajian:2012ys,Adhikari:2016bei} or models with an additional U(1) gauge symmetry \cite{Chen:2008yi,Choi:2020tqp,Choi:2020nan,Choi:2020udy}. {Recently, the authors of Ref. \citep{Choi:2021uhy} engineered a model in the context of supergravity that explicitely reproduces the kind of late 2-body decays considered here, and also provides a natural explanation for the small mass splitting that seems to be favoured by cosmic data.}

Decays to electromagnetically charged particles are tightly constrained by {\emph{Planck}} data~\cite{Slatyer:2016qyl,Poulin:2016anj}, $\gamma$-ray \cite{Cirelli:2012ut,Essig:2013goa} and cosmic-ray searches \cite{Jin:2013nta,Giesen:2015ufa}, typically requiring $\Gamma^{-1}\!>\!{\cal O}(10^{26}$s), with some level of  dependence on the decay channel. 
Still, a purely gravitational constraints, although weaker, is very interesting in the spirit of being `model-independent', while applying to models with decay to a dark sector, or to (non-interacting) neutrinos. The canonical example is perhaps that of the keV majoron \cite{Berezinsky:1993fm,Lattanzi:2008ds} decaying into relativistic neutrinos. Models of CDM decays with massive daughters have also been invoked as a potential solution to the observational discrepancies with CDM on small (sub-galactic) scales after structure formation
(e.g.~\cite{LinEtAl2001,SanchezSalcedo2003,CembranosEtAl2005,Kaplinghat2005,StrigariEtAl2007c,BorzumatiEtAl2008,PeterEtAl2010a,PeterEtAl2010,Choi:2020nan}).
Even more recently, decaying dark matter models were proposed \cite{KannikeEtAl2020,Xu:2020qsy,Choi:2020udy,Buch:2020xyt} as a way to explain the excess of events in the electronic recoils reported by the Xenon1T collaboration \cite{Xenon1tEtAl2020}.  
In the literature, most studies restricted themselves to massless daughter particles \cite{Audren:2014bca,Enqvist:2015ara,Berezhiani:2015yta,Poulin:2016anj,Nygaard:2020sow}, with the benefit of greatly simplifying the cosmological analysis, but limiting the true `model-independence' of the bound, and therefore its robustness. 
Nevertheless, some studies have attempted at including the effect of massive daughters in a cosmological context but either neglected cosmological perturbations of the daughter particles \cite{vattis_late_2019,Clark:2020miy,Haridasu:2020xaa} or were limited by computational power to perform a complete analysis against a host of cosmological data \cite{Aoyama:2011ba,Aoyama:2014tga}.

It has originally been suggested that DM decaying into massless daughters could help with cosmological tensions \cite{Enqvist:2015ara,Berezhiani:2015yta}, but careful analysis of this scenario in light of {\emph{Planck}} 2015 data has excluded this possibility \cite{Chudaykin:2016yfk,Poulin:2016nat} (although see Ref.~\cite{Bringmann:2018jpr} for a different conclusion if the decay rate is not constant). Attempting to go beyond these studies, the authors of Ref.~\cite{vattis_late_2019} suggested that considering a non-zero mass for (at least one of) the decay product would affect the phenomenology and allow for a resolution of the Hubble tension. 
However, a recent study based on a combination of both BAO and uncalibrated SNIa data-sets has been carried out in Ref.~\cite{Haridasu:2020xaa}. As opposed to Ref.~\cite{vattis_late_2019}, they conclude that a $\Lambda$DDM scenario does not predict higher $H_0$ values. 
This is in good agreement with model-independent analyses existing in the literature in which it has been established that a combination of BAO and uncalibrated SNIa data strongly constrain any late-time modification as a resolution to the Hubble tension (see,~e.g, Refs.~\cite{Poulin:2018zxs,Knox:2019rjx}).
A similar conclusion is also reached when CMB data are considered~\cite{Clark:2020miy}. Yet, these recent analyses were limited to the study of the effects of $\Lambda$DDM on the background evolution of the universe. 

In this paper, we perform a thorough analysis of the $\Lambda$DDM model in light of  up-to-date low- and high-redshift data-sets, including the effects of linear perturbations. 
To that end,  we introduce a new approximation scheme that allows to accurately and quickly compute the dynamics of the WDM linear perturbations by treating the WDM species as a viscous fluid.
In a companion paper \cite{Abellan:2020pmw}, thanks to this new fluid approximation, we have shown that the $\Lambda$DDM, while unable to ease the Hubble tension, can fully explain the low-$S_8$ measurement from recent weak lensing surveys. The aim of this paper is to: i) introduce the cosmological formalism of the $\Lambda$DDM model and a new approximation scheme that we developed to accurately describe linear perturbations of the warm daughter;  ii) discuss the background and perturbation dynamics of the $\Lambda$DDM model and its impact on the CMB and linear matter power spectrum; iii) compare the constraints obtained with the inclusion of perturbations to those obtained when neglecting them (as was done in the past literature); iv) test the robustness of the $\Lambda$DDM resolution to the $S_8-$tension to a number of changes in the analysis (different $S_8$ priors, different CMB datasets, marginalization over the lensing information in~{\emph{Planck}}, including constraints from the Xenon1T experiment on the model).

This work is structured as follows: in Section~\ref{sec_dcdm:model} we introduce the formalism and the novel approximation scheme; in Section~\ref{sec_dcdm:observables} we illustrate the $\Lambda$DDM impact on the relevant cosmological observables; in Section~\ref{sec_dcdm:analysis} we discuss the results of our data analyses; {in Section~\ref{sec:cmbs4} we show that a next generation CMB experiment (CMB-S4) can detect DDM}; finally, in Section~\ref{sec_dcdm:conclusion} we draw our conclusions.

\section{\label{sec_dcdm:model} Formalism of 2-body $\Lambda$DDM }

Hereafter, we adopt the Boltzmann formalism by Ref.~\cite{Aoyama:2014tga}, where the time-evolution of the Phase-Space Distribution (PSD) for both the mother and the daughter particles were derived. However, in Section~\ref{sec_dcdm:background} we explicitly show that, at the background level, such formalism is equivalent to the one by \cite{blackadder_dark_2014}.

While the (cold) parent particle can be safely described as a perfect fluid, computing the density perturbation evolution for the daughter particles requires a more sophisticated treatment. 
The central role in the game is played by $\ell_{\rm max} $, i.e. the highest multiple to consider when drawing up the hierarchy of equations describing the PSD of the daughter particles. 
In the massless case, the degrees of freedom associated to momentum can be removed after the PSD multipole decomposition~\cite{Poulin:2016nat}. 
Due to its non-trivial energy-momentum relation, this approach is not possible for the warm daughter particle. One has to study the full PSD evolution, which would be computationally prohibitive when performing MCMC analyses. 
For this reason, in Section~\ref{sec_dcdm:perturbation} {we provide a detailed description of a novel approximation scheme, devised in Ref.~\cite{Abellan:2020pmw}}, based on describing the WDM component as a viscous fluid on sub-Hubble scales.
This allows us to integrate out the momentum degrees of freedom and  the hierarchy of equations to be truncated at $\ell_{\rm max}=1$. 
We will show that the new, computationally faster scheme is accurate enough to be used for cosmological analyses, allowing to establish accurate and robust CMB limits on this class of models.

Our framework is characterized by two additional free parameters with respect to $\Lambda$CDM: the DCDM lifetime, $\Gamma^{-1}$, and the fraction of DCDM rest mass energy converted into DR, defined as follows \cite{blackadder_dark_2014}:
\begin{equation}\label{epsilon}
\eps = \frac{1}{2} \left(1-\frac{m^2_\wdm}{m^2_\dcdm}\right),
\end{equation}
where $0 \leq \eps \leq 1/2$. The lower limit corresponds to the standard CDM case, so that $\Omega_{\rm{cdm}}=\Omega_{\rm{dcdm}}+\Omega_{\rm{wdm}}$, 
whereas $\eps = 1/2$ corresponds to DM decaying solely into DR.
In general, small $\varepsilon$ values (i.e.~heavy massive daughters) and small $\Gamma$ values (i.e.~lifetimes much longer than the age of the universe) induce little departures from $\Lambda$CDM. 

Let us decompose the PSD function of the $j$-th dark component into a background contribution $\bar{f}_j$ plus a linear perturbation $\Delta f_j$ as
\begin{equation}\label{f_j}
f_j (k, q, \mu, \tau) = \bar{f}_j (q, \tau)+\Delta f_j (k, q, \mu, \tau),
\end{equation}
where $\tau$ is the conformal time\footnote{We use dots to indicate derivatives with respect to conformal time} and $j=$ $\{\dcdm, \dr, \wdm \}$.
The mean energy density and pressure $\bar{\rho}_j$ are obtained by integrating the background PSD, i.e.
\begin{align}
\bar{\rho_j} &= \frac{1}{a^4} \int_0^{\infty} dq \ 4\pi q^2 \mathcal{E}_j \bar{f}_j, \label{general_rho}  \\
\bar{P}_j  &= \frac{1}{3 a^4} \int_0^{\infty} dq \ 4\pi q^2 \frac{q^2}{\mathcal{E}_j}  \bar{f}_j \label{general_P},
\end{align}
where $\mathcal{E}_j = \sqrt{m_j^2 a^2+q^2}$ is the comoving energy of $j$-th species.
The linear perturbation term $\Delta f_j$ is generally expanded over Legendre polynomials:
\begin{equation}
\Delta f_j (k, q, \mu, \tau) = \sum_{\ell=0}^{\infty} (-i)^\ell (2\ell+1) \Delta f_{j,\ell} (k,q, \tau) P_\ell(\mu). \label{Pl}
\end{equation}
The perturbed energy density, pressure, energy flux and shear stress are thus given by:
\begin{align}
\delta \rho_j =\bar{\rho_j} \delta_j &= \frac{4\pi}{a^4} \int_0^{\infty} dq  q^2 \mathcal{E}_j \Delta f_{j,0}, \label{deltaj}  \\
\delta P_j = \bar{\rho}_j \Pi_j &= \frac{4\pi}{3 a^4} \int_0^{\infty} dq  q^2 \frac{q^2}{\mathcal{E}_j}  \Delta f_{j,0}, \label{pij} \\ 
(\bar{\rho_j}+\bar{P}_j )\theta_j &= \frac{4\pi k}{a^4} \int_0^{\infty} dq  q^2 q \Delta f_{j,1} \label{thetaj} , \\  
(\bar{\rho_j}+\bar{P}_j )\sigma_j &= \frac{8\pi}{3 a^4} \int_0^{\infty} dq  q^2 \frac{q^2}{\mathcal{E}_j}  \Delta f_{j,2}. \label{sigma}
\end{align}
We choose to work in the synchronous gauge, co-moving with the parent particle, that means that each $\Delta f_{\dcdm,\ell}$ vanishes except for the term $\Delta f_{\dcdm,0}$. 
In addition, since the parent particle is non-relativistic, the only relevant dynamical DCDM variables are $\bar{\rho}_\dcdm$ and $\delta_\dcdm$.
We follow the standard convention of referring to the scalar metric perturbations in this gauge as $h$ and $\eta$. 
We implement the equations describing the $\Lambda$DDM model in a modified version of the Boltzmann code \texttt{CLASS}\footnote{\url{http://class-code.net/}}\cite{Blas_2011,Lesgourgues:2011rh}. Our code is available at \url{https://github.com/PoulinV/class_majoron}.

Throughout the rest of this paper and unless stated otherwise,  we compare $\Lambda$DDM models at fixed $\omega_\dcdm^{\rm ini} \equiv \Odcdm h^2$ with $\Lambda$CDM models having the  same $\omega_{\rm cdm} $. All other cosmological parameters are fixed to $\{H_0 =67.7 \ \rm{km}/\rm{s}/\rm{Mpc}$, $\omega_{\rm cdm}=\omega_\dcdm^{\rm ini}  = 0.1194$, $\omega_{\rm b} = 0.0224$, $n_s=0.9673$, $\text{ln}(10^{10} A_s)=3.052$, $\tau_{\rm reio}=0.0582\}$, which constitutes our baseline $\Lambda$CDM model. 
These values correspond to the best-fit from the combined analysis (including $S_8$ data from weak lensing) of Ref.~\cite{Abellan:2020pmw}, which are however very similar to the $\Lambda$CDM best-fit parameters from {\emph{Planck}} 2018~\cite{Aghanim:2018eyx}. 

\subsection{\label{sec_dcdm:background}Background dynamics}
The background evolution of the mother and daughter particles is described at the phase-space level by their Boltzmann equations~\cite{Aoyama:2014tga}:
\begin{align}
\dot{\bar{f}}_\dcdm &= -a\Gamma \bar{f}_\dcdm, \label{f_dot_dcdm} \\
 \dot{\bar{f}}_\dr =  \dot{\bar{f}}_\wdm &=  \frac{a \Gamma \bar{N}_\dcdm (\tau)}{4 \pi q^2} \delta (q-a p_{\rm max}) \label{f_dot_dr}.
\end{align}
$\bar{N}_\dcdm = (\Odcdm \rho_{c,0} /m_\dcdm) e^{-\Gamma t}$ is the DCDM (comoving) number density\footnote{$\Omega^0_\dcdm \equiv \Odcdm  e^{-\Gamma t_0} $; $\Odcdm$ is the initial DCDM abundance, $t_0$ the age of the universe.}; $p_{\rm{max}}= |\vec{p}_\wdm|=|\vec{p}_\dr|  = m_\wdm \eps /\sqrt{1-2\eps}$ is the initial momentum of the daughter particles. 
 
By combining Eqs. \eqref{general_rho} and \eqref{general_P} with Eqs. \eqref{f_dot_dcdm} and \eqref{f_dot_dr} we obtain:
\begin{align}
\dot{\bar{\rho}}_\dcdm &= -3 \mathcal{H} \bar{\rho}_\dcdm -a \Gamma \bar{\rho}_\dcdm, \label{rho_dot_dcdm}\\
\dot{\bar{\rho}}_\dr &= -4 \mathcal{H} \bar{\rho}_\dr +\eps a \Gamma \bar{\rho}_\dcdm \label{rho_dot_dr}, \\
\dot{\bar{\rho}}_\wdm &= -3 (1+w) \mathcal{H} \bar{\rho}_\wdm  + (1-\varepsilon) a \Gamma  \bar{\rho}_\dcdm \label{rho_dot_wdm}.
\end{align}
$\mathcal{H} \equiv \dot{a}/a$ is the conformal Hubble parameter, and $w (\tau) \equiv \bar{P}_\wdm /\bar{\rho}_\wdm$ is the WDM Equation of State (EoS)\footnote{Notice that this expression does not coincide with the EoS used in \cite{vattis_late_2019}}.\ 

Eq.~\eqref{rho_dot_wdm} will be useful to analytically derive the fluid equations that we present in Section \ref{sec_dcdm:perturbation2}, but for a numerical resolution it is much simpler to use an integral formula for $\bar{\rho}_\wdm$, as it was done in Ref.~\cite{blackadder_dark_2014}. This formula can be obtained by integrating Eq.~\eqref{f_dot_dr} firstly with respect to $\tau$, and then with respect to $q$. 
The first integration requires using the relation $\delta (q - a p_{\rm max}) = \delta (\tau- \tau_q)/q \mathcal{H}$, where $\tau_q$ represents the conformal time when daughter particles with co-moving momentum $q$ are born, $q = a (\tau_q) p_{\rm{max}}$ \cite{Aoyama:2014tga}. For the second integral, changing the integration variable from $q$ to $a_q = a(\tau_q)$ leads to
\begin{align}
\bar{\rho}_\wdm (a) &=  \frac{  C }{a^4}\int_{0}^{a} da_q \frac{e^{-\Gamma t_q } }{\mathcal{H}_q}  \sqrt{ \eps^2 a_q^2+(1-2\eps )a^2 }, \label{rhowdm}
\end{align}
where $C \equiv \rho_{c,0}  \Odcdm \Gamma  $, $\mathcal{H}_q \equiv \mathcal{H} (a_q)$, $t_q \equiv t(a_q)$. 
Note the equivalence between Eq.~\eqref{rhowdm} and the analogous expression derived in Ref.~\cite{blackadder_dark_2014} with a different formalism.
Concerning the massless DR species, we simply take the limit $\varepsilon \rightarrow 1/2$ of Eq.~\eqref{rhowdm}. 

In both cases, the background evolution Eq.~\eqref{rhowdm} needs to be solved iteratively, as the Hubble parameter $\mathcal{H}$ depends on $\bar{\rho}_\wdm$ and $\bar{\rho}_\dr$  through the Friedmann equation. 
For a flat universe, this equation reads
\begin{equation}\label{Ha}
\mathcal{H}^2(a) = \frac{8 \pi G a^2}{3} \sum_i \bar{\rho}_i (a),
\end{equation}
where
\begin{align}
\sum_i \bar{\rho}_i (a) &= \bar{\rho}_{\rm{dcdm}}(a)+ \bar{\rho}_{\rm{dr}}(a)+ \bar{\rho}_{\rm{wdm}}(a) \nonumber \\ &+\bar{\rho}_{\gamma}(a)+\bar{\rho}_{\nu}(a)+\bar{\rho}_b(a)+\bar{\rho}_{\Lambda}.
\end{align}

Here $\bar{\rho}_{\gamma}$, $\bar{\rho}_{\nu}$, $\bar{\rho}_b$ and $\bar{\rho}_{\Lambda}$ denote the mean densities of photons, neutrinos, baryons and dark energy, respectively.  

\begin{figure}
    \centering
    \includegraphics[scale=0.45]{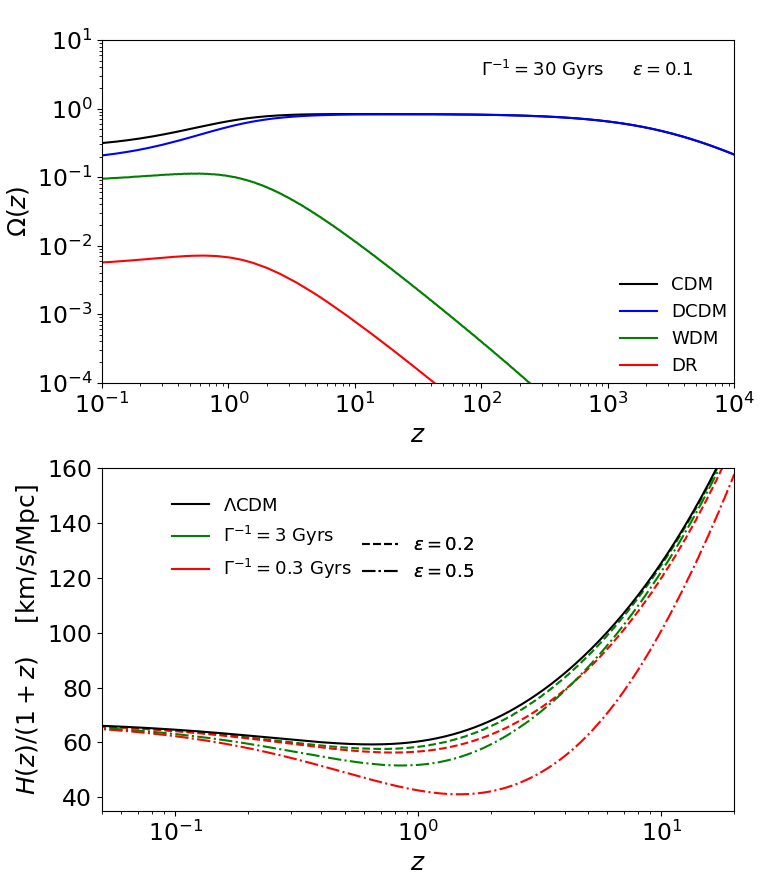}
    \caption{{\em Upper} $-$ Redshift evolution of the abundances of the DCDM, WDM and DR species, assuming $\Gamma^{-1}= 30 \ \text{Gyrs}$ and $\varepsilon=0.1$. We also show the abundance for a standard CDM species with $\Omega^0_{\rm cdm}=\Odcdm$. {\em Lower} $-$ Redshift evolution of the Hubble parameter for the $\Lambda$CDM and four different $\Lambda$DDM models. The Hubble parameter today is fixed to $H_0 =67.7 \ \rm{km}/\rm{s}/\rm{Mpc}$.  
    }
    \label{fig:background_dcdm}
\end{figure}

We show the evolution of the Hubble parameter for $\Lambda$CDM and several  $\Lambda$DDM models in lower panel of Fig. \ref{fig:background_dcdm}.  
In the upper panel of Fig. \ref{fig:background_dcdm}, we also show the evolution of the fractional densities  $\Omega_{\dcdm}(z)$, $\Omega_{\wdm}(z)$ and $\Omega_{\dr}(z)$ for a particular $\Lambda$DDM model ($\Gamma^{-1}= 30 \ \text{Gyrs}$ and $\varepsilon=0.1$) as well as $\Omega_{\rm cdm}(z)$ with the same initial amount of dark matter, namely $\Omega^0_{\rm cdm}=\Odcdm$.
The two new parameters affect $\mathcal{H}(a)$ as follows: at fixed $\eps$, a shorter lifetime $\Gamma^{-1}$ implies a lower 
Hubble parameter. This is clearly visible at $z\gtrsim1$. The behavior below $z\sim 1$ is due to our assumption of a flat universe and the requirement of fixing $H_0$:  to fulfill the budget equation, a lower DM abundance requires a larger $\Omega_{\Lambda}$, meaning that the period of accelerated expansion begins earlier with respect to $\Lambda$CDM. 
At fixed lifetime $\Gamma^{-1}$, a larger $\eps$ also induces a lower $\mathcal{H}(a)$, because more energy is converted into radiation, which dilutes faster. We thus anticipate a negative correlation between $\Gamma^{-1}$ and $\eps$ at the background level. The degeneracy can be  captured by the matter contribution from the WDM species, $\Omega_\wdm^{\rm m} \equiv \Omega_\wdm^0 (1-3w)$, which is well constrained by data. This quantity is roughly approximated by 
\begin{equation}
\label{eq:approx}
\Omega_\wdm^{\rm m} \simeq \Odcdm (1-e^{-\Gamma t_0}) \sqrt{1-2\varepsilon},
\end{equation}
 For small $\Gamma$ and small $\varepsilon$, $\Omega_\wdm^{\rm m} \propto \Gamma( 1-\varepsilon )$ and we expect data to constrain a parameter combination of the form $\varepsilon \propto \Gamma^{-1} $.

\subsection{\label{sec_dcdm:perturbation}Linear perturbation equations}

The continuity equation for the DCDM density perturbation can be obtained either by using the covariant conservation of the stress-energy tensor $T_{\mu \nu}$, or by integrating the corresponding first order Boltzmann equation (see Eq.~(2.20) in Ref.~\cite{Aoyama:2014tga}). 
As in the case of standard CDM, the resulting expression is
\begin{equation}
\dot{\delta}_\dcdm = -\frac{\dot{h}}{2}.
\end{equation}
To obtain the dynamical equations for the daughter particles, we combine Eq.~\eqref{Pl} with the corresponding first order Boltzmann equation (see Eq.~(2.31) in Ref.~\cite{Aoyama:2014tga}), so that we have the following hierarchy of equations:

\begin{align}
\frac{\partial \left(\Delta f_{j,0} \right) }{\partial \tau} &= -\frac{q k}{\mathcal{E}_j } \Delta f_{j,1} 
 + q\frac{\partial \bar{f}_j}{\partial q}  \frac{\dot{h}}{6}
+\dot{\bar{f}}_j \delta_\dcdm, \label{delta_f_0}\\ 
\frac{\partial  \left(\Delta f_{j,1} \right)}{\partial \tau} &= \frac{q k}{3\mathcal{E}_j } \left[\Delta f_{j,0}-2 \Delta f_{j,2} \right], \label{delta_f_1} \\
\frac{\partial \left(\Delta f_{j,2} \right) }{\partial \tau} &= \frac{q k}{5\mathcal{E}_j } \left[ 2 \Delta f_{j,1}-3 \Delta f_{j,3} \right] 
 - q\frac{\partial \bar{f}_j}{\partial q} \frac{(\dot{h}+6\dot{\eta})}{15},   \  \label{delta_f_2} \\
\frac{\partial \left(\Delta f_{j,\ell} \right) }{\partial \tau} &= \frac{q k}{(2l+1)\mathcal{E}_j } \left[ l \Delta f_{j,\ell-1}-(\ell+1) \Delta f_{j,\ell+1} \right] \nonumber \\
&\hspace{50mm} (\ell \geq 3), \label{delta_f_3}
\end{align}

where $j = \{\dr$, $\wdm \}$.  
The system of Eqs.~\eqref{delta_f_0}-\eqref{delta_f_3} is in the same form as the one for massive neutrinos \cite{Ma:1995ey}, except for the last term in Eq.~\eqref{delta_f_0} and the fact that the partial derivative $\partial \bar{f}_j / \partial q$ in Eqs.~\eqref{delta_f_0} and \eqref{delta_f_2} is now time-dependent.\

Given that the DR species satisfies the condition $q/\mathcal{E}_\dr =1$, the hierarchy of equations can be simplified by taking the moments
\begin{equation}
F_{\dr,\ell} \equiv \frac{1}{\rho_{c,0}} \int_0^{\infty} dq \ 4 \pi q^2 q \Delta f_{\dr,\ell}
\end{equation}
of Eqs.~\eqref{delta_f_0}-\eqref{delta_f_3}, and integrating over all the momentum degrees of freedom, so that
\begin{align}
\dot{F}_{\dr,0}   &= - k F_{\dr,1} 
 -\frac{2}{3} r_{\rm dr} \dot{h}+\dot{r}_{\rm dr} \delta_\dcdm, \label{F_0_dr} \\ 
\dot{F}_{\dr.1}   &= \frac{ k}{3}F_{\dr,0}-\frac{2k}{3} F_{\dr.2}, \label{F_1_dr}  \\ 
 \dot{F}_{\dr,2}  &= \frac{2 k}{5} F_{\dr,1}-\frac{3 k}{5} F_{\dr,3}
 + \frac{4}{15}r_{\rm dr} (\dot{h}+6\dot{\eta}),  \label{F_2_dr}  \\ 
 \dot{F}_{\dr,\ell} &= \frac{ k}{(2\ell+1)} \left[ \ell F_{\dr,\ell-1}-(\ell+1) F_{\dr,\ell+1} \right] \ \ \ \ \  \ (\ell \geq 3). \label{F_3_dr} 
\end{align}

We have adopted the convention $r_\dr \equiv a^4 \bar{\rho}_\dr /\rho_{c,0}$, as in Ref.~\cite{Poulin:2016nat}, which in the 2-body decay scenario under study leads to:
\begin{equation}
\dot{r}_\dr = \varepsilon a \Gamma 
(\bar{\rho}_\dcdm/\bar{\rho}_\dr) r_\dr. 
\end{equation}
The first three multipoles are given by:
\begin{equation}
F_{\dr,0} = r_\dr \delta_\dr, \hspace{2mm} F_{\dr,1} =\frac{4 r_\dr}{3 k} \theta_\dr, \hspace{2mm} F_{\dr,2} = 2 \sigma_\dr r_\dr. 
\end{equation}
We choose the maximum multipole $\ell_{\rm max}$ to truncate the hierarchies of equations according to the scheme proposed in Ref.~\cite{Ma:1995ey} for both massless and massive neutrinos, i.e.

\begin{align}
 \dot{F}_{\dr,\ell_{\rm max}} &= k F_{\dr,\ell_{\rm max}-1}-\frac{\ell_{\rm max}+1}{\tau} F_{\dr,\ell_{\rm max}}, \label{F_dr_max} \\    
 \Delta \dot{f}_{\wdm,\ell_{\rm max}} &= \frac{q k \Delta f_{\wdm,\ell_{\rm max}-1}}{\mathcal{E}_\wdm} -\frac{\ell_{\rm max}+1}{\tau} \Delta f_{\wdm,\ell_{\rm max}} . \label{delta_f_max}
\end{align}

\subsection{\label{sec_dcdm:perturbation2}A novel WDM fluid approximation scheme}

In order to compute the WDM dynamics one cannot integrate over the momentum degrees of freedom, as we did in the DR case. Indeed, when taking the moments of the hierarchy of Eqs.~\eqref{delta_f_0}-\eqref{delta_f_3}, higher velocity-weight integrals appearing at $\ell=2$ cannot be computed from the system of equations itself.

Therefore, one has to follow the evolution of the full time-dependent PSD 
to obtain the elements of the perturbed stress-energy tensor $\delta_\wdm$, $\theta_\wdm$ and $\sigma_\wdm$ through Eqs.~\eqref{deltaj}-\eqref{sigma}. 
A typical set-up for CMB analyses requires roughly $500$ wavenumbers, $50$ multipoles and $10^4$ momentum bins, i.e. $\mathcal{O}(10
^8)$ linear differential equations to be computed. On a single processor, this leads to runs with a CPU time of $1-2$ days per each parameter choice, making a systematic scan of the parameter space computationally prohibitive. 

To overcome the problem, we make use of a new fluid approximation for the WDM species, {introduced in Ref.~\cite{Abellan:2020pmw},} where the momentum dependence is removed, and one only needs to track the evolution of the first two multipoles. The number of linear differential equations to be solved is now reduced to $\mathcal{O}(10^3)$, with a CPU time per single run $\sim 30-40$ s. The novel approximation scheme is based on the treatment of massive neutrinos as a viscous fluid by Ref.~\cite{Lesgourgues:2011rh}, and it is only valid at scales deeply inside the Hubble radius, where high- and low-$\ell$ modes are effectively decoupled.
In App. \ref{sec_dcdm:appendix_numerics} we explicitly demonstrate the accuracy of this approximation.

Similarly to the DR case, the fluid equations can be derived by multiplying both sides of Eqs.~\eqref{delta_f_0} and \eqref{delta_f_1} by $4 \pi q^2 \mathcal{E}_\wdm a^{-4}$ and $4 \pi q^3 k a^{-4}$, respectively, and integrating over $q$. Then, by using Eqs.~\eqref{deltaj}-\eqref{sigma} and \eqref{rho_dot_wdm}, one can write down the continuity equation,
\begin{align}
\dot{\delta}_\wdm &= -3 \mathcal{H} (c^2_{\rm s}-w) \delta_\wdm -(1+w) \left(\theta_\wdm+\frac{\dot{h}}{2}\right) \nonumber \\
&+(1-\varepsilon) a\Gamma  \frac{\bar{\rho}_\dcdm}{\bar{\rho}_\wdm} (\delta_\dcdm-\delta_\wdm), \label{delta_dot_wdm}
\end{align}
and the Euler equation,
\begin{align}
\dot{\theta}_\wdm  &= -\mathcal{H} (1-3 c_g^2) \theta_\wdm + \frac{c^2_{\rm s}}{1+w}k^2\delta_\wdm -k^2 \sigma_\wdm \nonumber \\
&-(1-\varepsilon) a \Gamma  \frac{1+c_g^2}{1+w} \frac{\bar{\rho}_\dcdm}{\bar{\rho}_\wdm} \theta_\wdm .\label{theta_wdm}
\end{align}
Notice that we have introduced the WDM sound speed in the synchronous gauge, $c^2_{\rm s} \equiv \delta P_\wdm / \delta \rho_\wdm$, and the WDM adiabatic sound speed, $c_g^2 \equiv \dot{\bar{P}}_\wdm / \dot{\bar{\rho}}_\wdm$. 

The latter can be written as
\begin{equation}
c_g^2 = w (\dot{\bar{P}}_\wdm / \bar{P}_\wdm)(\dot{\bar{\rho}}_\wdm/ \bar{\rho}_\wdm )^{-1},
\end{equation}
and it can be computed as follows
\begin{align}
c_g^2 &= w \left(5- \frac{\mathfrak{p}_\wdm}{\bar{P}_\wdm} -\frac{\bar{\rho}_\dcdm}{\bar{\rho}_\wdm} \frac{a \Gamma}{3 w \mathcal{H}} \frac{\varepsilon^2}{1-\varepsilon} \right) \nonumber \\
 &\times\left[3(1+w)-\frac{\bar{\rho}_\dcdm}{\bar{\rho}_\wdm} \frac{a\Gamma}{\mathcal{H}} (1-\varepsilon) \right]^{-1} \label{sound_speed}.
\end{align}
Here $\mathfrak{p}_\wdm$ denotes the so-called pseudo-pressure, a higher momenta integral of $\bar{f}_\wdm$ which is reduced to the standard pressure in the relativistic limit \cite{Lesgourgues:2011rh}. 

Obtaining an analytical expression for $c^2_{\rm s}$ is less straightforward, since we do not have a dynamical equation for the pressure perturbation $\delta P_{\rm wdm}$. 
In Ref.~\cite{Lesgourgues:2011rh} it is assumed that $c^2_{\rm s}$ is scale-independent and approximately equal to $c_g^2 $. 
For the WDM species, we have found that this assumption leads to accurate results for the CMB power spectrum, but not for the matter power spectrum. 
In fact, calculations using the full Boltzmann hierarchy of Eqs.~\eqref{delta_f_0}-\eqref{delta_f_3} reveal that $c_{\rm s}^2 $ exhibits a particular $k$-dependence that cannot be captured by a background quantity such as $c_g^2$. 
In particular, $c_{\rm s}^2 $ gets slightly enhanced on scales $k > k_{\rm fs}$, where $k_{\rm fs}$ is the free-streaming scale of the WDM species, defined as
\begin{equation}
k_{\rm fs} (\tau) \equiv \sqrt{\frac{3}{2}} \frac{\mathcal{H}(\tau)}{c_g(\tau)}.
\label{free-streaming}
\end{equation}
It is possible to gain a semi-analytic understanding of this behaviour by building a formal equation for the evolution of $c^2_{\rm s}$, as detailed in App. \ref{sec_dcdm:appendix_cs2}.
To account for such an enhancement, we adopt the following prescription
\begin{eqnarray}
c_{\rm s}^2(k, \tau) = c_g^2(\tau) \left[ 1+(1-2\varepsilon) T(k/k_{\rm fs}) \right],
\label{sync_sound_speed}
\end{eqnarray}
where the function $T(x) = 0.2\sqrt{x}$ has been fitted to the sound speed obtained using the full Boltzmann hierarchy, for the parameter values $\varepsilon=0.5, 0.1, 0.01, 0.001$ and $\Gamma/H_0 = 0.1, 1, 10$. The factor $(1-2\varepsilon)$ is inserted to make the $k$-dependent correction vanishingly small close to the relativistic limit, where $c^2_{\rm s} \simeq c_g^2 \simeq 1/3$.

In order to trace the evolution of the shear $\sigma_\wdm$ one could follow a similar approach to that of Ref.~\cite{Lesgourgues:2011rh}, where the authors obtained a dynamical equation for the neutrino shear by means of an improved truncation scheme at $\ell_{\rm max}=2$. 
We tested the implementation of a generalization of that equation suitable to the decaying case, but we found it to be only relevant close to the relativistic case $\varepsilon \simeq 1/2$, when it reduces to the DR shear equation from Ref.~\cite{Enqvist:2015ara}.
In this regime, the dynamics of the daughter particles do not significantly impact the CMB and matter power spectra, we thus decided to not include any dynamical equation for the shear of the WDM species, when switching to the fluid approximation.
In practice, we simply set $\sigma_\wdm$ to a constant value, obtained via integration of the second PSD multipole in the Boltzmann hierarchy (see Eq.~\eqref{sigma}).
We explicitly checked that this approach yields better results rather than simply setting $\sigma_\wdm =0$ in the fluid equations, or than using the DR shear equation of Ref.~\cite{Enqvist:2015ara} when $\varepsilon=1/2$

Finally, let us recall that, strictly speaking, these equations are valid in the synchronous gauge co-moving with the DCDM. 
In practice, however, the synchronous gauge in \texttt{CLASS} \cite{Blas_2011} is coded with respect to the CDM.  
Nevertheless, for adiabatic initial conditions, one can choose $\theta_{\rm ini,dcdm} = \theta_{\rm ini, cdm} = 0$, such that the gauge co-moving with CDM is also co-moving with DCDM at all times~\cite{Audren:2014bca}. 
Hence, hereinafter we neglect this irrelevant complication.

\subsection{\label{sec_dcdm:effect_perturbations}Dynamics of perturbations}

\begin{figure}
    \centering
    \includegraphics[scale=0.5]{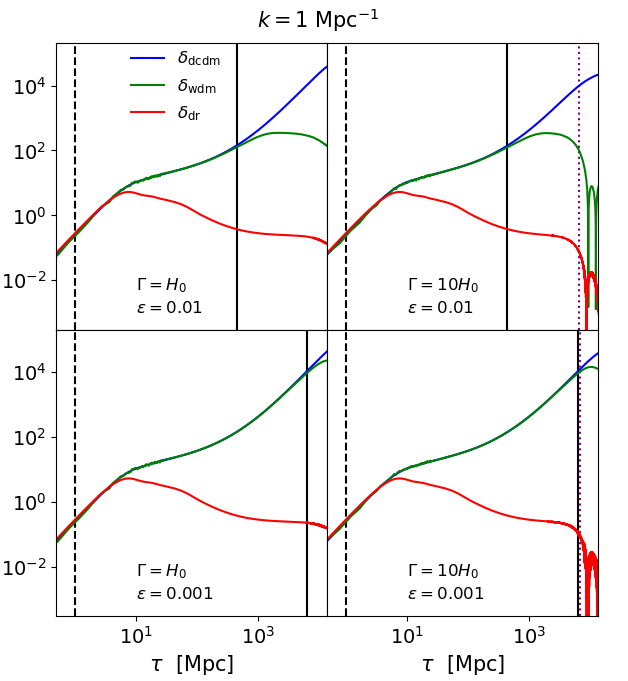}
    \caption{Time evolution of the linear density perturbations of the DCDM, WDM and DR species, corresponding to a wavenumber $k= 1 \ \text{Mpc}^{-1}$. Each panel displays the perturbations for a different combination of the parameters $\Gamma$ and $\varepsilon$. The black dashed and solid lines indicate the times of horizon crossing (at $\mathcal{H}(\tau)=k$) and WDM free-streaming scale crossing (at $k_{\rm fs}(\tau)=k$), respectively. The purple dotted line indicates the characteristic decay time, given by $t(\tau)=\Gamma^{-1}$.}
    \label{fig:perts_k3}
\end{figure}

Before discussing the signatures of varying the parameters $\Gamma$ and $\varepsilon$ on the relevant cosmological observables, it is worth having a look at the linear perturbations of the mother and daughter particles. 
In Fig. \ref{fig:perts_k3} we show the evolution of the linear density perturbations for the DCDM, WDM and DR species, corresponding to a mode that enters the horizon very early (i.e.,~$k= 1 \ \rm{Mpc}^{-1}$). To clarify the impact of the two extra free parameters, the perturbations are shown for DCDM lifetimes similar and smaller than the age of the universe ($\Gamma^{-1} = H_0^{-1} \sim 14.5 \ \text{Gyrs}$ and $\Gamma^{-1} = (10 H_0)^{-1} \sim 1.5 \ \text{Gyrs}$) , and for massive daughters behaving either as warm or cold particles ($\varepsilon =10^{-2}$ and $\varepsilon = 10^{-3}$). \ 

One can see that initially, the perturbations of the daughter species always track those of the mother, because the coupling term dominates the dynamics (i.e., the ratios $\rho_\dcdm/\rho_\wdm$ and $\rho_\dcdm/\rho_\dr$ are large). 
When a mode crosses the free-streaming scale, the pressure support of the daughter particles becomes important and the perturbations develop oscillatory features. 
For the DR species, the free-streaming scale simply corresponds to the horizon ($k_{\rm fs} \sim \mathcal{H}$,) while for the WDM it corresponds to a larger value, given by $k_{\rm fs} \sim \mathcal{H}/c_g$ (see Eq.~\eqref{free-streaming}). 
This time- and scale-dependent power suppression, together with the different background dynamics, lead to key signatures on the CMB and matter power spectra, as we discuss later.
By comparing upper and lower panels of Fig.~\ref{fig:perts_k3}, it is clear that the value of the WDM free-streaming scale is essentially determined by the value of $\varepsilon$, as expected.
On the other hand, by comparing left and right panels, one notices that the intensity of the oscillations due to the pressure support can be compensated by the coupling to the DCDM if the lifetime is long enough, as the daughter particles keep being produced. \

\begin{figure}
    \centering
    \includegraphics[scale=0.5]{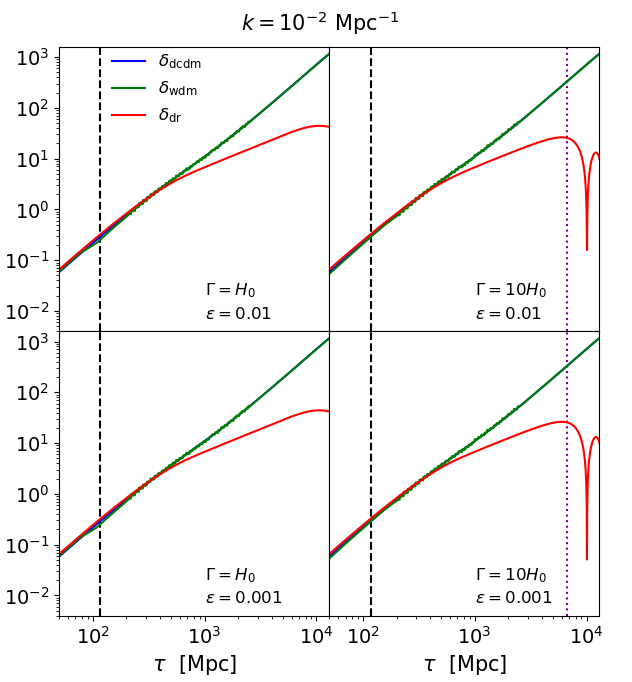}
    \caption{Same as in Fig. \ref{fig:perts_k3}, but corresponding to a wavenumber $k = 10^{-2} \ \text{Mpc}^{-1}$. In this case, the WDM perturbations never cross the free-streaming scale. }
    \label{fig:perts_k1}
\end{figure}

Interestingly, the decoupling time of the daughter perturbations from the mother perturbations is always set by the free-streaming crossing time, and not by the characteristic decay time. In order to illustrate that, in Fig. \ref{fig:perts_k1} we show perturbations corresponding to a smaller wave-number, $k = 10^{-2} \ \text{Mpc}^{-1}$, that enters the horizon much later. For this mode, the WDM species does not have time to cross the free-streaming length (the crossing will occur in the future), so that $\delta_\dcdm$ and $\delta_\wdm$ remain equal, even if the lifetime is smaller than the age of the universe. This can be understood from the fluid Eqs. \eqref{delta_dot_wdm}-\eqref{theta_wdm}: when the decay term that includes $\Gamma$ is relevant, $\delta_{\wdm}$ is driven by $\delta_{\rm dcdm}$. Therefore, the WDM density perturbation $\delta_\wdm$ will continue to track the behaviour of $\delta_\dcdm$, as long as the free-streaming scale is not crossed, i.e., as long as the pressure term, containing $c_s^2 k^2 \delta_\wdm$, is small compared to $\ddot{\delta}_\wdm \sim \delta_\wdm \mathcal{H}^2$.

\section{\label{sec_dcdm:observables} Observable impact of Dark Matter decays with warm daughters}

\subsection{\label{sec_dcdm:impact_pk}The linear matter power spectrum}

\begin{figure*}
\includegraphics[scale=0.28]{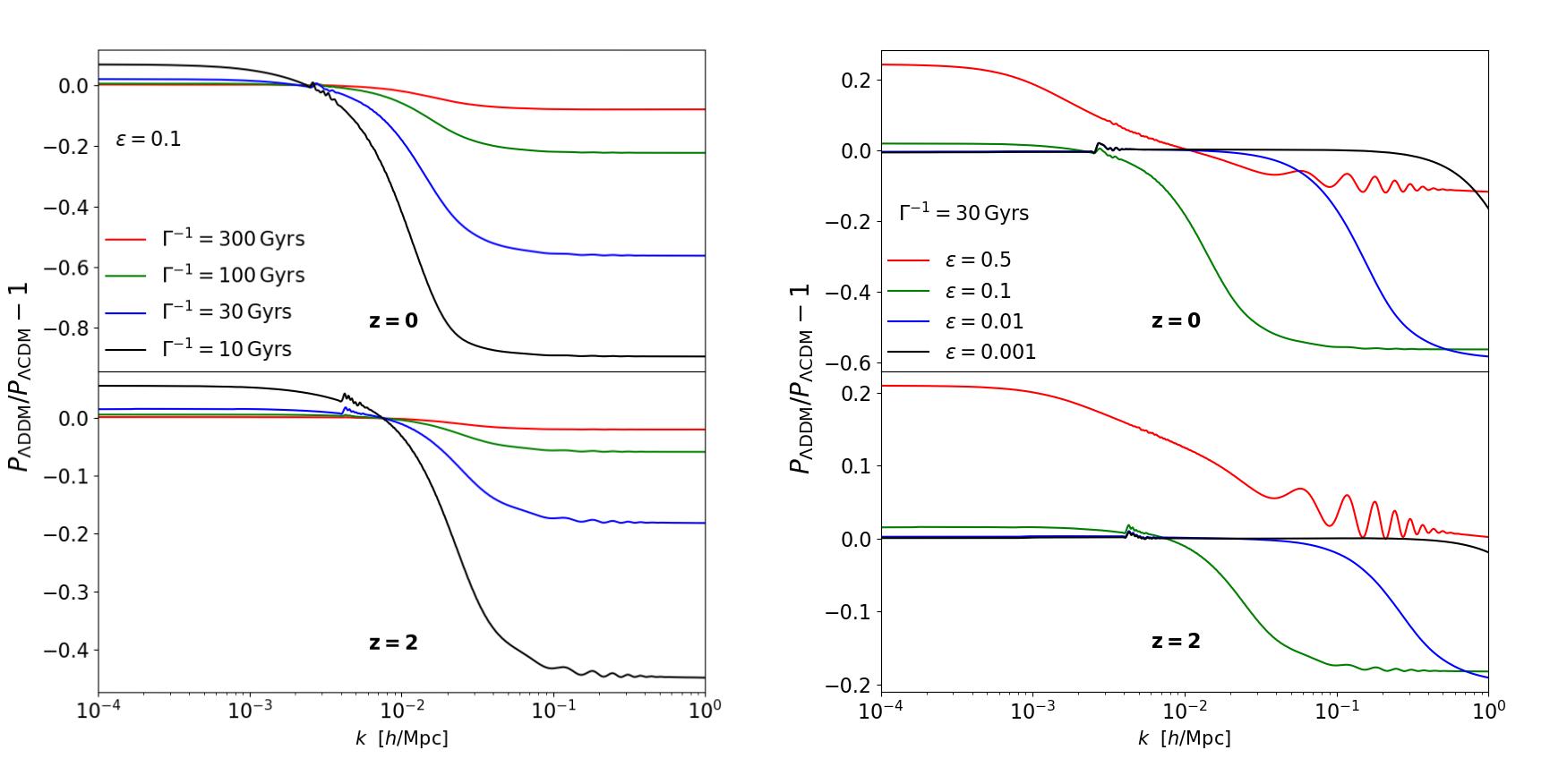}
\caption{ {\em Left} $-$ Residuals of the linear matter power spectrum at $z=0$ (upper) and $z=2$ (lower) for several values of the lifetime $\Gamma^{-1} = 10, 30, 100, 300 \ \text{Gyrs}$ and a fixed DR energy fraction $\varepsilon=0.1$. Residuals are taken with respect our baseline $\Lambda$CDM model. {\em Right} $-$ Same as in the left, but for several DR energy fractions $\varepsilon=0.5, 0.1, 0.01, 0.001$ and a fixed lifetime $\Gamma^{-1}=30$ Gyrs.}
\label{fig:pk_suppression}
\end{figure*}

Firstly, we focus on describing the effects of the 2-body decay on the linear matter power spectrum, since this will allow to better understand some of the effects on the CMB spectra. In this Section, we use the same $\Lambda$CDM parameters as in Section~\ref{sec_dcdm:background}, except for fixing $100\theta_s = 1.04217$ instead of $H_0$, to better connect with CMB observations, that accurately pin down $\theta_s$. In Fig.~\ref{fig:pk_suppression}, we compare the residual differences in the linear power spectra (at redshifts $z=0$ and $z=2$) with respect to our baseline $\Lambda$CDM. The left panel shows several lifetimes and a fixed DR energy fraction $\varepsilon=0.1$, while the right panel shows a fixed lifetime $\Gamma^{-1}=30$ Gyrs and several values of the DR energy fraction $\varepsilon$.

 One important feature of the C+WDM scenarios, such  as the one considered in this work, is that they are expected to produce a suppression in the linear matter power spectrum at scales smaller than a `cutoff scale', with a non-trivial shape~\cite{Murgia:2017lwo,Murgia:2018now,Miller:2019pss,Bohr:2020yoe}. The cut-off scale is determined by the free-streaming scale of the WDM species, $k_{\rm fs}$, given by Eq.~\eqref{free-streaming}. On scales $k > k_{\rm fs}$, pressure becomes important and WDM particles cannot stay confined in gravitational potential wells, which inhibits structure formation.

Fig.~\ref{fig:pk_suppression} clearly illustrates that, while the parameter $\varepsilon$ fixes the value of the cut-off $k_{\rm fs}$\footnote{By looking at Eqs. \eqref{sound_speed} and \eqref{free-streaming}, we see that for small values of  $\varepsilon$, the cut-off approximately satisfies the scalings $k_{\rm fs} \propto c_g^{-1} \propto w^{-1/2} \propto \varepsilon^{-1}$.}, the lifetime 
$\Gamma^{-1}$ essentially determines the depth of the suppression at very small scales. This is to be expected, since the amount of power suppression grows with the WDM abundance, which increase for smaller lifetimes. In a similar way, the effects of decay become less important when considering the matter spectrum at a higher redshift $z=2$, since the abundance of WDM was smaller in the past. 

In general, for late-time decay scenarios (well after recombination) as the one studied in this work, it is possible to distinguish three different regimes depending on the value of $\varepsilon$, as it is shown in the right panel of Fig.~\ref{fig:pk_suppression}:
 
\begin{itemize}
    
    \item \textbf{Non-relativistic decay}: if $\varepsilon \lesssim 0.001$ (black curve), the WDM leaves the expansion rate unaffected, since its contribution to the matter density, $\bar{\rho}_\wdm (1-3 \omega) \simeq \bar{\rho}_\wdm $, compensates the reduction in the DCDM density, $\bar{\rho}_{\rm dcdm}$. 
    In addition, the WDM free-streaming length is very small, inducing a power suppression at $k \gtrsim 1 \  h\mathrm{Mpc}^{-1}$. Such scales are beyond the range of scales probed by the observables considered in this work, so in this regime the WDM is almost degenerate with standard CDM.

    \item \textbf{Relativistic decay}: if $\varepsilon \simeq 0.5 $ (red curve), the WDM component acts as DR, which can appreciably reduce the expansion rate\footnote{Note that since now we are fixing $100 \theta_s$ instead of $H_0$, the Hubble rate $H(z)$ can increase with respect to $\Lambda$CDM at $z\lesssim1$, once dark energy starts to dominate. However, this effect is small for long lifetimes, and at early times the Hubble rate is still smaller than in $\Lambda$CDM. }. Moreover, the free-streaming length $k_{\rm fs}$ gets as large as the horizon, so that the WDM does not cluster at all. The reduction in the Hubble friction is balanced by a reduction in the clustering density of the daughter particles, $\delta \rho_\wdm \simeq \delta \rho_\dr \simeq 0$, inducing a very little overall suppression of the growth of fluctuations.  However, there is an impact coming from another background effect: the significant decrease in the co-moving matter density, $a^3 \rho_{\rm m}$, leading in turn to an increase of the angular diameter distance, thereby a reduction of $\theta_s$. This effect can be compensated by increasing $H_0$, which shifts the location of the peak, $k_{\rm eq}/(a_0 H_0)$, towards smaller $k$ in the matter power spectrum -- since we are keeping the matter-radiation equality era fixed. The net effect on the residuals is twofold, a large-scale enhancement and small-scale suppression of power.

    \item \textbf{Warm decay}: For intermediate values of $\varepsilon$, namely $ 0.001 \lesssim  \varepsilon \lesssim 0.5$ (green and blue curves), the WDM component partially contributes to the matter energy density, leaving to an expansion rate almost unchanged. However, the values of $k_{\rm fs}$ that determine the cut-off scale in the matter power spectrum are not as small as in the case of non-relativistic decay, leading to $\delta \rho_{\wdm} \simeq 0$ for $k>k_{\rm fs}$. Hence, on scales $k>k_{\rm fs}$ the Hubble friction gets enhanced with respect to the clustering density $\delta \rho_{\rm tot}$, slowing down the growth of DCDM perturbations. The net impact on the matter power spectrum is thus a suppression on intermediate scales, somewhat similar to that induced by massive active neutrinos~\cite{Lesgourgues:2006nd,Giusarma:2016phn,Parimbelli:2018yzv}.

\end{itemize}

\subsection{\label{sec_dcdm:impact_cmb}The CMB angular power spectrum}

\begin{figure*}
\includegraphics[scale=0.23]{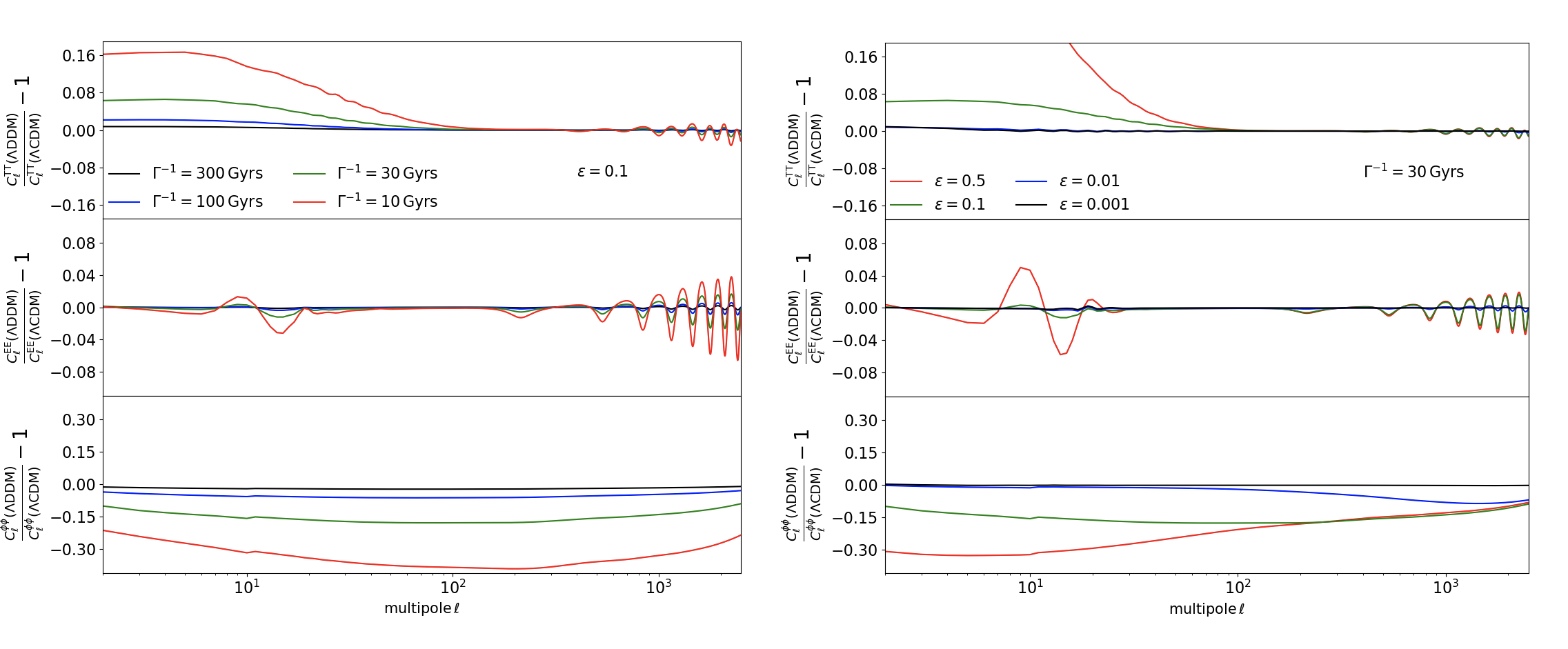}
\caption{ {\em Left} $-$ Residuals (with respect our baseline $\Lambda$CDM model) of the CMB lensed TT (upper), EE (middle) and lensing potential (lower) power spectra for several values of the lifetime $\Gamma^{-1} = 10, 30, 100, 300 \ \text{Gyrs}$ and a fixed DR energy fraction $\varepsilon = 0.1$. {\em Right} $-$ Same as in the left, but for several values of the DR energy fraction $\varepsilon =0.5, 0.1, 0.01, 0.001$ and a fixed lifetime $\Gamma^{-1} = 30 \ \text{Gyrs}$.  }
\label{fig:dcdm_vs_lcdm} 
\end{figure*}

We now discuss the impact of the 2-body decay scenario on the CMB anisotropy temperature and polarization angular power spectra, as well as on the lensing potential power spectrum reconstructed from the CMB 4-point correlation function. 
In the left panel of Fig.~\ref{fig:dcdm_vs_lcdm} we report the residuals of the (lensed) TT, EE and lensing potential power spectra with respect to our baseline $\Lambda$CDM, for different lifetimes $\Gamma^{-1}$ and a fixed DR energy fraction $\varepsilon = 0.1$. The effects, more and more pronounced as the lifetime decreases, can be understood as follows: 
\begin{itemize}
    
    \item  At the background level, the decay decreases the value of $\Omega_{\rm m}$ with respect to $\Lambda$CDM. This is compensated by an increase in $\Omega_{\Lambda}$ (earlier beginning of $\Lambda$-domination) and thus an enhancement in the Late Integrated Sachs-Wolfe (LISW) effect, leaving a signature in the low-$\ell$ TT power spectrum. Furthermore, a modified background history alters quantities integrated along $z$, such as $\tau_{\rm reio}$, which impacts the multipoles $\ell \sim 10$ in the EE power spectrum.

    \item At the perturbation level, the late-time reduction of $a^3\bar{\rho}_\dcdm$ implies a reduction of the quantity $a^2\delta\rho_\dcdm$, which acts as a source of gravity through the Poisson equation. This induces a damping in the metric fluctuations, and hence yields a further enhancement of the LISW effect. Furthermore, the suppression in the matter power spectrum and in $\Omega_{\rm m}$ lowers the amplitude of the lensing potential power spectrum, consequently reducing the smoothing of the peaks in the high-$\ell$ part of both the TT and EE spectra, as one can see from the `wiggles' in the corresponding plots. 
    
\end{itemize}

In the right panel of Fig.~\ref{fig:dcdm_vs_lcdm} we show the CMB residuals for a fixed $\Gamma^{-1}$ and various values of $\varepsilon$. The effects can be readily understood:

\begin{itemize}
\item At the background level, smaller values of $\varepsilon$ weaken the effects previously discussed, because the decay product dilute in a way similar to dark matter. Namely, the decrease in $\Omega_{\rm m}$ is less prominent due to the significant WDM contribution (i.e., the increase in $\Omega_{\Lambda}$ is shallower), and the impact on $\tau_{\rm reio}$ is smaller. Therefore, the signatures in the low-$\ell$ part of the TT and EE spectra become less visible.

\item At the level of perturbations, $\varepsilon$ leads to some interesting signatures on the LISW effect and on the lensing potential. Since the LISW effect is only relevant for small multipoles $\ell$ (i.e., very large scales), one just needs to look at wavenumbers such that $k < k_{\rm fs}$. On these scales, the growth suppression does not play any role, and the decrease in $a^2\delta\rho_\dcdm$ due to the decay gets partially compensated by the increase in $a^2 \delta \rho_\wdm$, which is more significant for smaller values of $\varepsilon$. Thus, the damping in the metric fluctuations is less relevant for smaller $\varepsilon$, reducing the LISW enhancement. 

\item Regarding the effects on the lensing potential, one can see that the suppression in the corresponding power spectrum monotonically decreases for smaller $\varepsilon$. Naively, one might expect the opposite, since we have argued that the matter power suppression increases for small $\varepsilon$. This can be understood by looking at the CMB lensing kernel $W(z)$, given by the following expression \citep{Manzotti:2017oby}
\begin{equation}
\centering
W(z) = \frac{3 \Omega_{\rm m}}{2}\frac{H_0^2}{H(z)} (1+z)\chi (z) \frac{\chi_{\ast}-\chi(z)}{\chi_{\ast}},
\label{kernel}
\end{equation}
where $\chi_{\ast}$ is the co-moving distance to the last-scattering surface. Firstly, the CMB lensing kernel  peaks at $z \sim 2$, where the suppression is less important (see bottom panels of Fig.~\ref{fig:pk_suppression}). Secondly, it gets highly suppressed for higher values of $\varepsilon$, and this effect can dominate over the effect on the matter power spectrum. This suppression happens mainly due to the smaller $\Omega_{\rm m}$, as we have verified by computing $W(z)$ for several $\Lambda$DDM models, with and without including the factor $\Omega_{\rm m}$. Note that coincidentally, the effects on the lensing power spectrum at high $\ell$ are very similar for $\varepsilon=0.5$ and $\varepsilon=0.1$: this is because in the former case $\Omega_{\rm m}$ is highly reduced and the small scales power spectrum is almost unaffected, while the opposite occurs in the latter case. 
\end{itemize}

Let us finally remark that, even if the effects of varying either $\varepsilon$ or $\Gamma$ on the observables are different, one can easily exploit the degeneracy mentioned at the background level to get different couples of values (large $\Gamma$ and small $\varepsilon$ or vice-versa) with a similar cosmological signature, especially on the CMB. We do indeed expect the MCMC analysis to show a negative correlation in the reconstructed 2D posteriors for $\varepsilon$ and~$\Gamma$.

\section{\label{sec_dcdm:analysis}Data analysis}

\subsection{\label{sec_dcdm:data}Data and method}
 We now confront the $\Lambda$DDM cosmology to a host of recent cosmological observations. Our goal is to set constraints on the lifetime of DM  and the mass-ratio of the mother and daugther particles. We also wish to check to what extent the $\Lambda$DDM model can play a role in resolving cosmological tensions as claimed in the recent literature \cite{vattis_late_2019,Abellan:2020pmw}. 
 To that end, we perform comprehensive MCMC analyses with the \texttt{MontePython-v3}\footnote{\url{https://github.com/brinckmann/montepython_public}} \cite{Audren:2012wb,Brinckmann:2018cvx} code interfaced with our modified version of \texttt{CLASS}, considering various combinations of the following data-sets:
\begin{itemize}
    
    \item The BAO measurements from 6dFGS at $z=0.106$~\cite{Beutler:2011hx}, SDSS DR7 at $z=0.15$~\cite{Ross:2014qpa}, BOSS DR12 at $z=0.38, 0.51$ and $0.61$~\cite{Alam:2016hwk}, and the joint constraints from eBOSS DR14 Ly-$\alpha$ auto-correlation at $z=2.34$~\cite{Agathe:2019vsu} and cross-correlation at $z=2.35$~\cite{Blomqvist:2019rah}.
    
    \item The measurements of the growth function $f\sigma_8(z)$ (FS) from the CMASS and LOWZ galaxy samples of BOSS DR12 at $z = 0.38$, $0.51$, and $0.61$~\cite{Alam:2016hwk}.
    
    \item The Pantheon SNIa catalogue, spanning redshifts $0.01 < z < 2.3$~\cite{Scolnic:2017caz}.
    
    \item The low-$\ell$ CMB TT, EE, and the high-$\ell$ TT, TE, EE data\footnote{In our main analysis, we use the `lite' version of the {\sc clik} likelihood. We verify that this leads to negligible differences with respect to the full likelihood in App. \ref{sec_dcdm:appendix_planck}.} + the gravitational lensing potential reconstruction from {\emph{Planck}} 2018~\cite{Aghanim:2018eyx}.

    \item The KIDS-1000+BOSS+2dFLens~\cite{Heymans:2020gsg}, DES-Y1~\cite{Abbott:2017wau} and KIDS-1000+Viking+DES-Y1~\cite{Joudaki:2019pmv} weak lensing data, compressed as a a split-normal likelihood,~i.e., $S_8=0.766^{+0.02}_{-0.014}$, $S_8 = 0.773_{-0.02}^{+0.026}$, $S_8=0.755_{-0.021}^{+0.019}$, respectively.
    
    \item The local measurement of the Hubble constant from SH0ES\footnote{A new version of the SH0ES measurement \cite{Riess:2020fzl} was published during completion of this work. We do not expect it to have any impact on our conclusions.}, modelled with a Gaussian likelihood centered on $H_0 = 74.03 \pm 1.42$ km/s/Mpc~\cite{Riess:2019cxk}.
    
\end{itemize}

We start by performing two distinct sets of studies in order to illustrate the importance of taking CMB data into account when studying the $\Lambda$DDM scenarios, even in the long-lived regime (i.e.~when the DCDM decays after CMB decoupling):

\vspace{0.05cm}
(i) a background-only analysis against BAO\footnote{As discussed in Section~\ref{sec_dcdm:Intro}, we calibrate BAO data by imposing a Gaussian prior on the sound horizon at recombination $r_s (z_{\rm rec})=144.7 \pm 0.5\ \rm{Mpc}$, to not to spoil CMB data~\cite{Aghanim:2018eyx}.} and Pantheon SNIa data;

\vspace{0.05cm}
(ii) full analyses including linear perturbations, where we combine the data-set used in (i) with CMB TT, TE, EE + lensing data, with and without including the aforementioned informative priors on $S_8$ and $H_0$.

\vspace{0.05cm}

In the case  of (i), the parameter space is fully characterized by the following free parameters:

\begin{equation*}
\left\{ \Odcdm, H_0, \Gamma, \eps \right\},
\end{equation*}
whereas in (ii) the whole parameter space is described by:
\begin{equation*}
\left\{\Omega_b h^2, \text{ln} \left(10^{10} A_s\right), n_s, \tau_{\rm reio},  \Odcdm, H_0, \Gamma, \eps \right\}.
\end{equation*}

For both (i) and (ii) we adopt logarithmic priors on $\eps$ and $\Gamma$~\footnote{For comparisons with previous works, an useful conversion is the following  $\mathrm{Log}_{10} (\Gamma / \mathrm{Gyrs}^{-1}) \simeq \mathrm{Log}_{10} (\Gamma / \mathrm{km} \ \mathrm{s}^{-1} \mathrm{Mpc}^{-1})-2.991 $. }, namely,
\begin{equation*}
    \begin{tabular}{c}
        $ -4 \leq {\mathrm{Log}_{10}}~\eps \leq {\mathrm{Log}_{10}}(0.5)$,   \\
         $-4 \leq {\mathrm{Log}_{10}}~(\Gamma/{\rm Gyrs}^{-1}) \leq 1$,
    \end{tabular}
\end{equation*}
and a flat prior on the initial DCDM abundance:
\begin{equation*}
0 \leq \Odcdm \leq 1. 
\end{equation*}

 Secondly, we explore the possibility of resolving the infamous Hubble and $S_8$ tensions, and the `$A_{\rm lens}$' anomaly that exists within {\emph{Planck}} data.
We then test the robustness of our results to various changes in the pipeline, and in particular to trading the high-$\ell$ {\emph{Planck}} CMB data for those from the SPT collaboration \cite{Henning:2017nuy}, which are known to be less in tension with local $S_8$ measurements, {as well as ACTPol data \cite{Aiola:2020azj}, which shows a level of tension with $S_8$ measurements similar to Planck}. Finally, we briefly discuss the viability of the 2-body decay scenario as solution for the Xenon1T anomaly~\cite{Xenon1tEtAl2020}.

We {adopt flat priors on all other parameters}, and we set two massless and one massive active neutrino species with $m_{\nu} = 0.06 \ \text{eV}$, following {\emph{Planck}}'s conventions~\cite{Aghanim:2018eyx}. 
We assume our MCMC chains to be converged when the Gelman-Rubin criterion $R-1 < 0.02$ \cite{Gelman:1992zz}. 
To extract the best-fit parameters, we make use of the {\sc Minuit} algorithm \cite{James:1975dr} through the {\sc iMinuit} python package\footnote{\url{https://iminuit.readthedocs.io/}}.
In App.~\ref{sec_dcdm:chi2} we report all individual $\chi^2$'s per each of the analyses performed.

\subsection{\label{sec_dcdm:results} General constraints: background vs. linear perturbations}

\begin{figure*}
\includegraphics[scale=0.5]{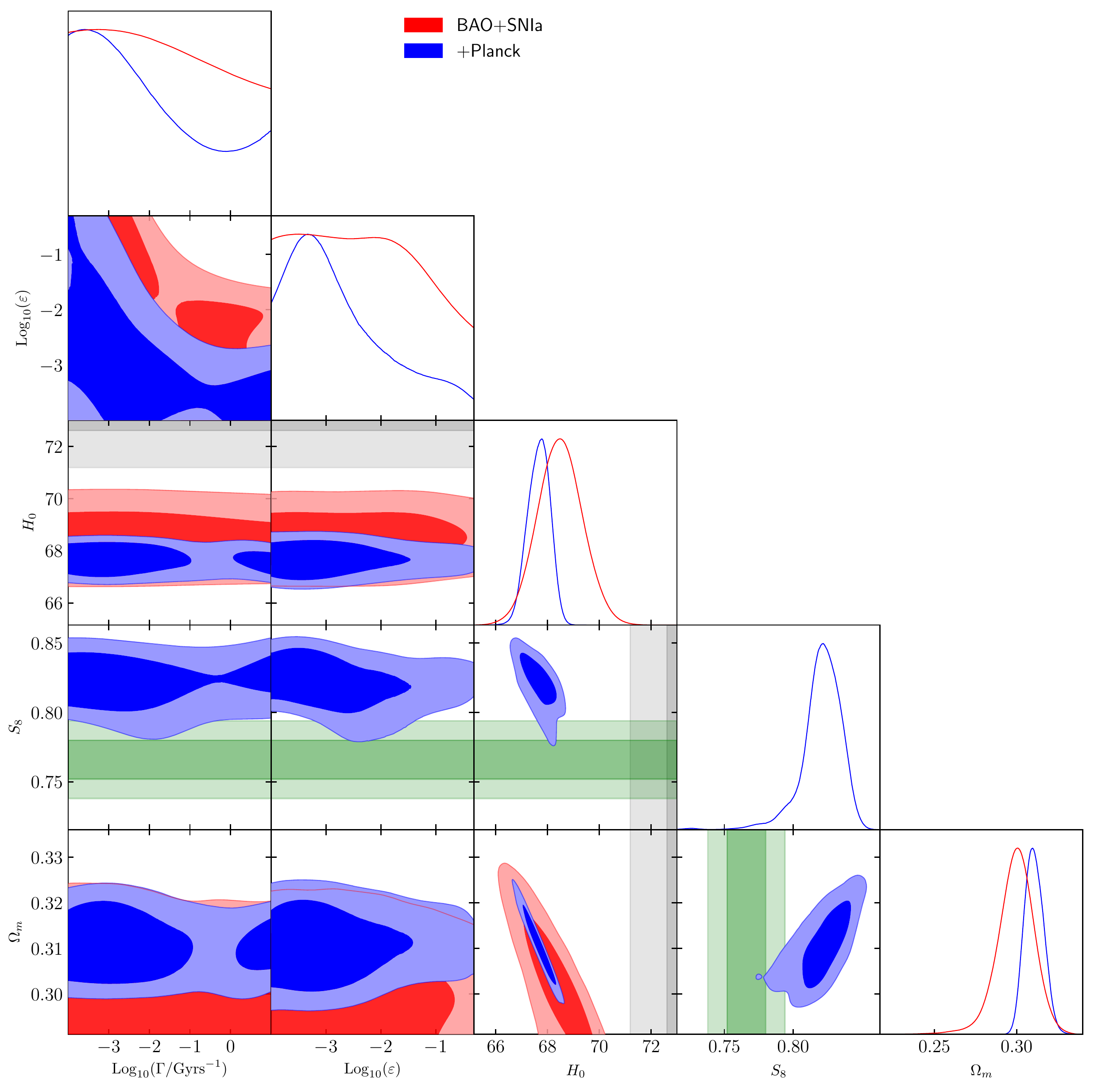}
\caption{\label{fig:Prior} 1 and 2D marginalized posterior distributions for the cosmological parameters relevant for our analysis. Hereafter, unless otherwise stated, the green shaded bands refer to the joint $S_8$ measurement from KiDS-1000+BOSS+2dFLens, {while the gray bands stand for the $H_0$ measurement by the SH0ES collaboration. Note that the BAO+SNIa analysis is based on background evolution only, whereas the BAO+SNIa+CMB analysis also includes linear perturbations (see the main text for further details)}.}
\end{figure*}

In Fig.~\ref{fig:Prior} we show the 2$\sigma$ posterior distributions of the parameters characterizing the $\Lambda$DDM model,  ${\mathrm{Log}_{10}}(\eps)$ and $ {\mathrm{Log}_{10}}(\Gamma/{\rm Gyrs}^{-1})$.
Red contours refer to the background-only analysis (i), while blue contours refer to the full analysis (ii), at the perturbation level, when CMB data are also added. 
From BAO+SNIa and {\emph{Planck}} data, the DDM is not detected. 
We confirm our expectation from Sec.~\ref{sec_dcdm:background} and \ref{sec_dcdm:observables} that there is a negative correlation between ${\mathrm{Log}_{10}}(\eps)$ and $ {\mathrm{Log}_{10}}(\Gamma/{\rm Gyrs}^{-1})$, apparent within all data sets: as $\varepsilon$ decreases, the 95\% C.L. upper limit on $\Gamma$ relaxes. In practice, we find that for decay rates $\Gamma \sim 10^{-3}-10^{-1} \text{Gyrs}^{-1}$, our $2 \sigma$ exclusion curve is roughly described by $\varepsilon \simeq 1.6 \times 10^{-4} (\Gamma / \text{Gyrs}^{-1})^{-1.1}$. 
For large $\Gamma$ and small $\varepsilon$, the factor `$1-e^{-\Gamma t_0}$' in Eq.~\ref{eq:approx} reaches 1 faster than `$\sqrt{1-2\varepsilon}$' (assuming $t_0 \simeq 13.8$ Gyrs), explaining why our constraint on the DR energy fraction becomes flat even for very large decay rates. 
In particular, for small decay rates  $ {\mathrm{Log}_{10}}(\Gamma/{\rm Gyrs}^{-1})\lesssim -3$ and very massive daughters ${\mathrm{Log}_{10}}(\eps)\lesssim-2.7$ the DCDM behaves like CDM, leading to departures indistinguishable from $\Lambda$CDM.  
Of utmost importance, we find that constraints on the $\Lambda$DDM free parameters become much stronger when CMB data are included, increasing by more than one order of magnitude over the whole parameter space, contrarily to the naive expectations that CMB data do not weigh-in on late-time decays.

Interestingly, we see that the background-only analysis predicts a value of $H_0$ slightly higher than the one inferred assumed $\Lambda$CDM, {though it can be noticed that the standard $\Lambda$CDM value for $H_0$ is still perfectly compatible, due to larger uncertainties with respect to the full analysis, which anyhow} pulls $H_0$ back to its standard value. This suggests that this kind of models is not suitable for relieving the Hubble tension, contrarily to earlier claims \cite{vattis_late_2019}, and in agreement with Refs.~\cite{Haridasu:2020xaa,Clark:2020miy}. However, in contrast to Ref.~\cite{Clark:2020miy},  we observe a significant decrease in the $S_8$ contours for $\varepsilon \sim 0.01$ and $\Gamma^{-1} \sim 10^2$ Gyrs. We attribute this disagreement to the fact that Ref.~\cite{Clark:2020miy} does not include a treatment of WDM perturbations, which are responsible for the suppression in the matter power spectrum. Hence, the 2-body decay presented here could potentially reconcile the inferred value of $S_8$ with its direct measurements from LSS observations, as pointed out in Ref.~\cite{Abellan:2020pmw}. We present an explicit comparison of our constraints with those from Ref.~\cite{Clark:2020miy} in App.~\ref{sec_dcdm:appendix_comparison}.

Finally, in order to compare our constraints on $\Gamma$ with previous literature, we have carried out a MCMC analysis including BAO + SNIa + \emph{Planck} data, but fixing\footnote{This is not equivalent to directly reading the constraints on $\Gamma$ at $\varepsilon = 0.5$ from the $\Gamma$ vs.~$\varepsilon$ contours, since the 95\% C.L. derived from a $\chi^2$ distribution with different degrees of freedom correspond to different $\Delta\chi^2$.} $\varepsilon = 0.5$  (i.e. in the limit in which the daughter particle behaves as dark radiation). We find a  $2\sigma$ upper limit on the DCDM decay rate of $\text{log}_{10} (\Gamma/[\text{Gyr}^{-1}]) \lesssim -2.67 $, corresponding to $\Gamma^{-1} \gtrsim 468 \ \rm{Gyrs}$. { Our constraints on DM decays to DR are three times tighter than those found in works using older Planck data \cite{Audren:2014bca,Poulin:2016nat} but also $\sim 40\%$ tighter than Ref.~\cite{Nygaard:2020sow} due to the use of a logarithmic prior on $\Gamma$ as opposed to linear.}

\begin{table}
\scalebox{0.95}{
 \begin{tabular}{|l|c|c|} 
 \hline
 Parameter & BAO+SNIa & +\emph{Planck} \\ 
 \hline
$100 \ \omega_b $								& -- & $2.243(2.244)_{-0.013}^{+0.014}$ \\
$\Omega_{\rm dcdm}^{\rm ini}$ 					& $0.2529(0.2532)_{-0.01}^{+0.0098}$ & $0.2606(0.2619)_{-0.0054}^{+0.0051}$ \\
$ H_0 /[{\rm km/s/Mpc}]$ 						& $68.48(68.44)_{-0.91}^{+0.88}$ & $67.71(67.71)_{-0.43}^{+0.42}$ \\
$\text{ln}(10^{10} A_s)$						&  --  &  $3.051(3.052)_{-0.015}^{+0.014}$ \\
$n_s$											&  --  &  $0.9674(0.9672)_{-0.0038}^{+0.0038}$ \\
$\tau_{\rm reio}$								&  --  & $0.0576(0.0582)_{-0.0079}^{+0.0069}$ \\
$\text{log}_{10} (\Gamma/[{\rm Gyr}^{-1}])$		&   unconstrained(0.09) & unconstrained($-3.86$)  \\
$\text{log}_{10} (\varepsilon)$					&   unconstrained(-2.89)   & $-2.69(-2.97)_{-1.3}^{+0.32}$ \\
$\Omega_{\rm m}$								&  $0.299(0.2992)_{-0.011}^{+0.013}$  & $0.3102(0.3109)_{-0.0058}^{+0.0056}$ \\
$S_8$											&  --  & $0.821(0.828)_{-0.011}^{+0.017}$ \\
\hline
 $\chi^2_{\rm min}$ & 1036.6  & 2053.4 \\
 \hline
\end{tabular}}
\caption{The mean (best-fit) $\pm 1\sigma$ errors of the cosmological parameters from our $\Lambda$DDM analyses against BAO + SNIa and BAO + SNIa + \emph{Planck}. For each data-set, we also report the best-fit $\chi^2$.}
\label{tab:nos8}
\end{table}

\subsection{\label{sec_dcdm:tensions}Implications for cosmological tensions and Xenon1T}

\subsubsection{\label{sec_dcdm:h0}The $H_0$ tension}

In order to test the implications of the 2-body decay for cosmological tensions, we conduct a run that include the local measurement of $H_0$ from SH0ES~\cite{Riess:2019cxk}, CMB, BAO and SNIa data. For the sake of brevity we do not report the results of the runs here. We find that the shape of the posterior probabilities is almost unchanged, except for a tiny shift in $H_0$ to a higher value, $H_0 = 68.21\pm0.4 \  \rm{km}/\rm{s}/\rm{Mpc}$. We thus confirm the inability of this model to resolve the Hubble tension. This was expected since it had already been shown through a model independent reconstruction of the late-time dynamics of the dark sector that any late-time solution that does not modify the sound horizon at recombination is expected to fail when combining BAO with SNIa data (e.g.~\cite{Poulin:2018zxs,Lemos:2018smw,Knox:2019rjx,Benevento:2020fev}). {Although this is not of material importance in the context of the present work, let us note that recent Refs.~\cite{Camarena:2021jlr,Efstathiou:2021ocp} emphasized that a more correct way of combining Pantheon and SH0ES is through a prior on the intrinsic magnitude of SN1a. Yet, this does not affect our conclusions, as it was explicitly shown in Ref.~\cite{Schoneberg:2021qvd}}.

\subsubsection{\label{sec_dcdm:s8}The $S_8$ tension and the role of priors}

\begin{figure}
    \centering
    \includegraphics[scale=0.42]{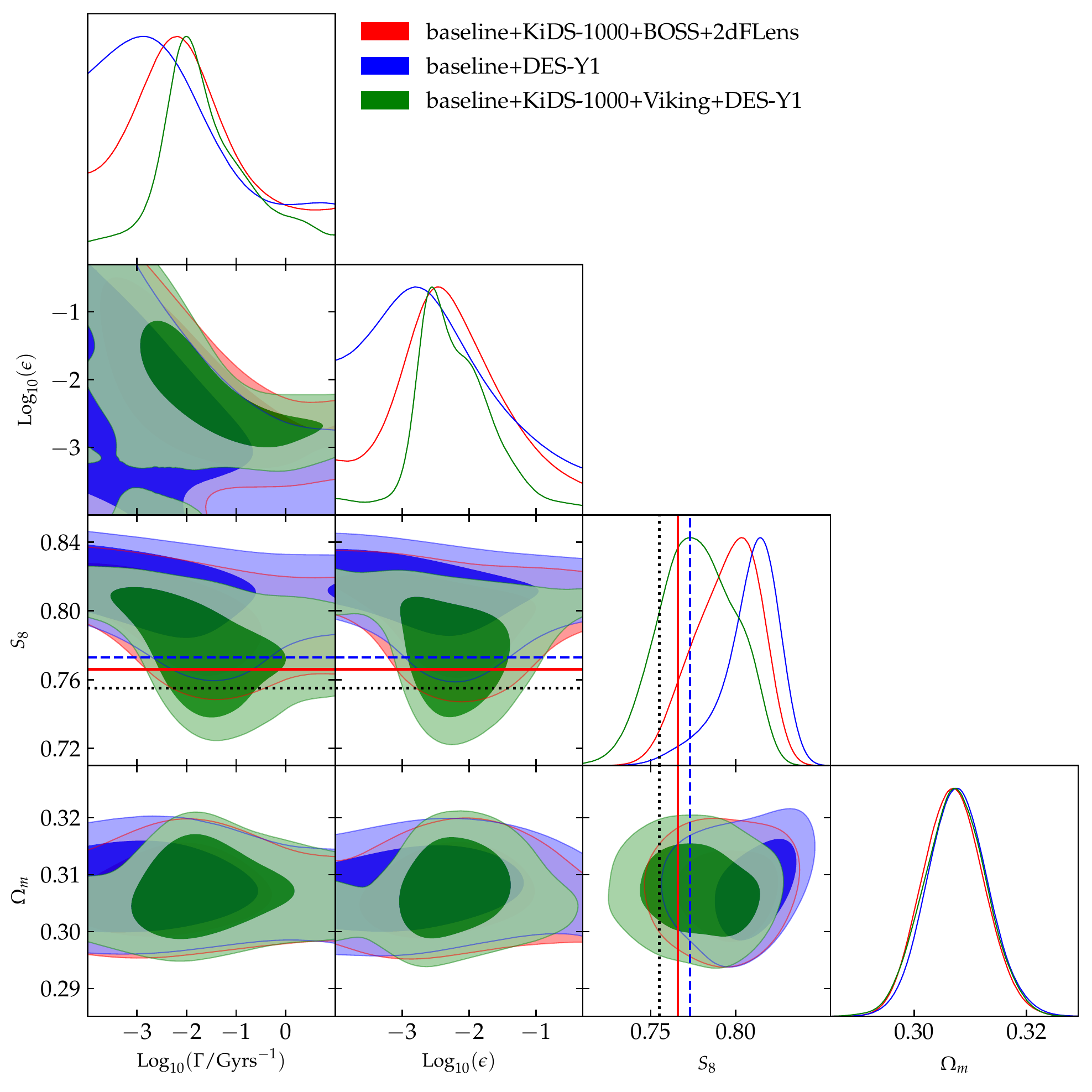}
    \caption{2D posterior distribution of a subset of parameters from analyses performed by imposing three different priors on $S_8$, based on the three sets of weak lensing measurements considered, whose mean values are shown as a solid, dashed and dotted line, for KiDS+BOSS+2dFLens, DES and KiDS+Viking+DES respectively. {Note that all the three analyses also include our baseline data-set,~i.e.~{\it Planck}}+BAO+SNIa.}
    \label{fig:s8}
\end{figure}

\begin{table*}
 \scalebox{1.0}{

\begin{tabular}{|l|c|c|c|} 
 \hline
 Parameter & w/ KiDS+BOSS+2dFLens & w/ DES & {w/ KiDS+Viking+DES} \\ 
 \hline
$100 \ \omega_b $                                    & $2.246(2.241) \pm 0.013$      & $2.245(2.241)_{-0.013}^{+0.014}$ & $2.246(2.244)_{-0.013}^{+0.014}$ \\
$\Omega_{\rm dcdm}^{\rm ini}$                        & $0.2581(0.2606)_{-0.0054}^{+0.005}$      & $0.2585(0.2610)_{-0.0051}^{+0.005}$ & $0.2585(0.2603)_{-0.0056}^{+0.0053}$ \\
$H_0 /[{\rm km/s/Mpc}]$                              & $67.92(67.70)_{-0.42}^{+0.43}$      & $67.88(67.66)_{-0.42}^{+0.41}$ & $67.89(67.73)_{-0.47}^{+0.42}$ \\
$\text{ln}(10^{10} A_s)$                             & $3.048(3.052)_{-0.016}^{+0.014}$      & $3.048(3.054)_{-0.015}^{+0.014}$ & $3.050(3.047)_{-0.015}^{+0.015}$ \\
$n_s$                                                & $0.9682(0.9673) \pm 0.0037 $      & $0.9681(0.9671)_{-0.0038}^{+0.0037}$ & $0.9681(0.9671)_{-0.004}^{+0.0036}$  \\
$\tau_{\rm reio}$								     & $0.0570(0.0582)_{-0.0077}^{+0.0071}$      & $0.0568(0.0590)_{-0.0078}^{+0.0069}$ &  $0.0578(0.0554)_{-0.0077}^{+0.007}$  \\
$\text{log}_{10} (\Gamma/[{\rm Gyr}^{-1}])$		     & $-1.89(-1.74)_{-1.5}^{+0.82}$      & $-2.15(-1.97)_{-1.8}^{+0.41}$        &  $-1.62(-1.08)_{-1}^{+0.81}$  \\
$\text{log}_{10} (\varepsilon)$					     & $-2.28(-2.16)_{-0.78}^{+0.8}$      & $< -2.14(-2.10)$        &  $-2.23(-2.52)_{-0.59}^{+0.48}$  \\
$\Omega_{\rm m}$								     & $0.3071(0.3099)_{-0.0057}^{+0.0053}$      & $0.3078(0.3107)_{-0.0054}^{+0.0055}$ &  $0.3073(0.3085)_{-0.0058}^{+0.0058}$  \\
$S_8$										         & $0.795(0.767)_{-0.016}^{+0.024}$      & $0.809(0.784)_{-0.01}^{+0.021}$      &  $0.778(0.763)_{-0.023}^{+0.025}$  \\
\hline
$\chi^2_{\rm min}$ & 2055.0   & 	2054.8	&  	2055.9  		\\
\hline
$\footnotesize{\chi^2_{\rm min} (\Lambda\text{DDM}) -\chi^2_{\rm min} (\Lambda\text{CDM})} $ & -5.7 & -2.3 & -8.6 \\
\hline
\end{tabular}}
\caption{The mean (best-fit) $\pm 1\sigma$ errors of the cosmological parameters from our BAO + SNIa + \emph{Planck} + $S_8$, being the latter from KiDS+BOSS+2dFLens, DES-Y1 and KiDS-1000+Viking+DES-Y1 analyses. For each data-set we also report its best-fit $\chi^2$, and the $\Delta\chi^2$ with respect to the analogous $\Lambda$CDM best-fit model.}
\label{tab:s8}
\end{table*}

{As illustrated in Sec.~\ref{sec_dcdm:results}, in a CMB+BAO+SNIa analysis of the $\Lambda$DDM scenario, not including weak lensing data, the $S_8$ tension is only marginally alleviated and DDM is not detected.}
On the other hand, in Ref.~\cite{Abellan:2020pmw} it was shown that, by including a $S_8$-prior as measured by KiDS+BOSS+2dFLens, the $\Lambda$DDM model can fully resolve the tension if DM decays with a lifetime of $\Gamma^{-1} \simeq  55 \ \text{Gyrs}$
and converts a fraction $\varepsilon\simeq 0.7 \ \%$ of its rest mass energy into kinetic energy for the massless component. The reason for this is explained in detail in the latter reference: it is due to the fact that the presence of WDM at late-times reduce the amplitude of the matter power spectrum on small scales, thereby decreasing $\sigma_8$, without changing significantly the value of $\Omega_m$. This results in a smaller $S_8$ without running into disagreement with measurements from $\Omega_m$ from BAO and SNIa data. Moreover, the fact that the decays occurs at late-times leaves the CMB power spectra largely unchanged. 

In this former study, a split-normal likelihood on $S_8$ as determined by KiDS-1000+BOSS+2dFLens~\cite{Heymans:2020gsg},~i.e.  $S_8=0.766^{+0.02}_{-0.014}$ was used. A more accurate approach would have been to include the full galaxy shear and clustering power spectra. 
Making use of the full likelihood would however require the ability to compute the matter power spectrum on non-linear scales in a $\Lambda$DDM universe, a task that 
is beyond the scope of this paper.
Let us note that it has been established in various cases that the reconstructed $S_8$ value only mildly vary from one model to another. In particular, the KiDS collaboration has established that the reconstructed value of $S_8$ is insensitive to the neutrino mass \cite{Hildebrandt:2018yau} -- a model that has physical effects very similar to the $\Lambda$DDM model. 
This provides confidence in making use of a  prior on $S_8$ derived in the $\Lambda$CDM context. 
Nevertheless, to highlight the impact of a different $S_8$ measurement, we replace the $S_8$ prior from KiDS-1000+BOSS+2dFLens by the one determined in the combined analysis KiDS+Viking+DES-Y1\footnote{Note that this combined analysis includes a photo-metric redshift correction applied to DES result, slightly lowering the $S_8$ value compared to what is advocated by the DES collaboration \cite{Abbott:2017wau}.} \cite{Joudaki:2019pmv}, $S_8=0.755_{-0.021}^{+0.019}$ and with the DES-Y1 data only \cite{Abbott:2017wau}, $S_8 = 0.773_{-0.02}^{+0.026}$. The results of these analyses are reported in Tab.~\ref{tab:s8}, and shown in Fig.~\ref{fig:s8}, where we directly compare the results of the three different runs\footnote{The mean $\pm 1\sigma$ errors, and the best-fit of the cosmological parameters in the KiDS-1000+BOSS+2dFLens analysis are reported in the first column of Tab.~\ref{tab:alens}}. From the reconstructed parameters

\begin{eqnarray}
\text{log}_{10} (\Gamma/[{\rm Gyr}^{-1}]) & = & -1.89_{-1.5}^{+0.82}~~~~~~{\rm KiDS1000}\nonumber\\
\text{log}_{10} (\varepsilon)& = &-2.28_{-0.78}^{+0.8}~~~~~~{\rm+BOSS\!+2dFLens}
\nonumber\\ 
S_8& = &0.795_{-0.016}^{+0.024},\nonumber\\
\nonumber\\
\text{log}_{10} (\Gamma/[{\rm Gyr}^{-1}])& = &-1.62_{-1}^{+0.81}~~~~~~{\rm KiDS\!+Viking}  \nonumber\\
\text{log}_{10} (\varepsilon)& = &-2.23_{-0.59}^{+0.48}~~~~~~{\rm +DES-Y1}\nonumber\\ 
S_8 & = & 0.778_{-0.023}^{+0.025} ,\nonumber\\
\nonumber\\
\text{log}_{10} (\Gamma/[{\rm Gyr}^{-1}])& = &-2.15_{-1.8}^{+0.41}~~~~~~~~{\rm DES-only}   \nonumber\\ 
\text{log}_{10} (\varepsilon)& = &-2.52_{-1.5}^{+0.38}\nonumber\\
S_8 & = & 0.809_{-0.01}^{+0.021},\nonumber
\end{eqnarray}
one can see that the level of preference is higher for the KiDS\!+Viking\!+DESY1 case while it is lower in the DES-only case.
In the former case, we nevertheless note a slight degradation of $\chi^2_{\rm min}\sim+1$ compared to the `baseline' analysis of Ref.~\cite{Abellan:2020pmw}, while in the latter case the $\chi^2_{\rm min}$ does not sensibly change.
This explicitly demonstrates that the statistical significance of the DDM ``detection'' is strongly driven by the level of tension of the $S_8$ value used in the analysis ({see the values of $\Delta \chi^2$ with respect to $\Lambda$CDM reported in Tab.~\ref{tab:s8}}). If the $S_8$ tension increases in the future, the preference for $\Lambda$DDM over $\Lambda$CDM would likely increase. On the other hand, if the $S_8$ tension disappears, cosmological data would {not favour the $\Lambda$DDM scenario compared to the standard $\Lambda$CDM scenario}.

\subsubsection{\label{sec_dcdm:s8_alens}The $A_{\rm lens}$ anomaly}

\begin{table}
 \scalebox{1.}{

\begin{tabular}{|l|c|} 
 \hline
 Parameter & w/ $A_{\rm lens}$ \\ 
 \hline
$100 \ \omega_b $                               & $2.262(2.260)_{-0.015}^{+0.016}$ \\
$\Omega_{\rm dcdm}^{\rm ini}$                   & $0.2506(0.2526)_{-0.0059}^{+0.0049}$ \\
$ H_0 /[{\rm km/s/Mpc}]$                        & $68.56(68.38)_{-0.45}^{+0.5}$ \\
$\text{ln}(10^{10} A_s)$                        & $3.025(3.032)_{-0.018}^{+0.02}$ \\
$n_s$                                           & $0.9725(0.9718)_{-0.0039}^{+0.0042}$ \\
$\tau_{\rm reio}$                               & $0.0474(0.0506)_{-0.008}^{+0.0098}$ \\
$\text{log}_{10} (\Gamma/[{\rm Gyr}^{-1}])$     & $-2.10(-1.49)_{-1.9}^{+0.39}$\\
$\text{log}_{10} (\varepsilon)$                 & unconstrained (-2.47)\\
$\Omega_{\rm m}$                                & $0.2991(0.3009)_{-0.0066}^{+0.0053}$ \\
$S_8$                                           & $0.784(0.768)_{-0.014}^{+0.018}$ \\
$A_{\rm lens}^{\rm TTTEEE}$                     & $1.208(1.192)_{-0.064}^{+0.066}$ \\
$A_{\rm lens}^{\phi\phi}$                       & $1.086(1.072)_{-0.041}^{+0.035}$ \\
\hline
 $\chi^2_{\rm min}$ & 2043.2\\
 \hline
\end{tabular}}
\caption{The mean (best-fit) $\pm 1\sigma$ errors of the cosmological parameters from our BAO + SNIa + \emph{Planck} + $S_8$ (from KiDS+BOSS+2dFLens) analysis performed by marginalizing over the amplitude of the lensing potential $A_{\rm lens}$. We also report the best-fit $\chi^2$.}
\label{tab:alens}
\end{table}

\begin{figure}
    \centering
    \includegraphics[scale=0.35]{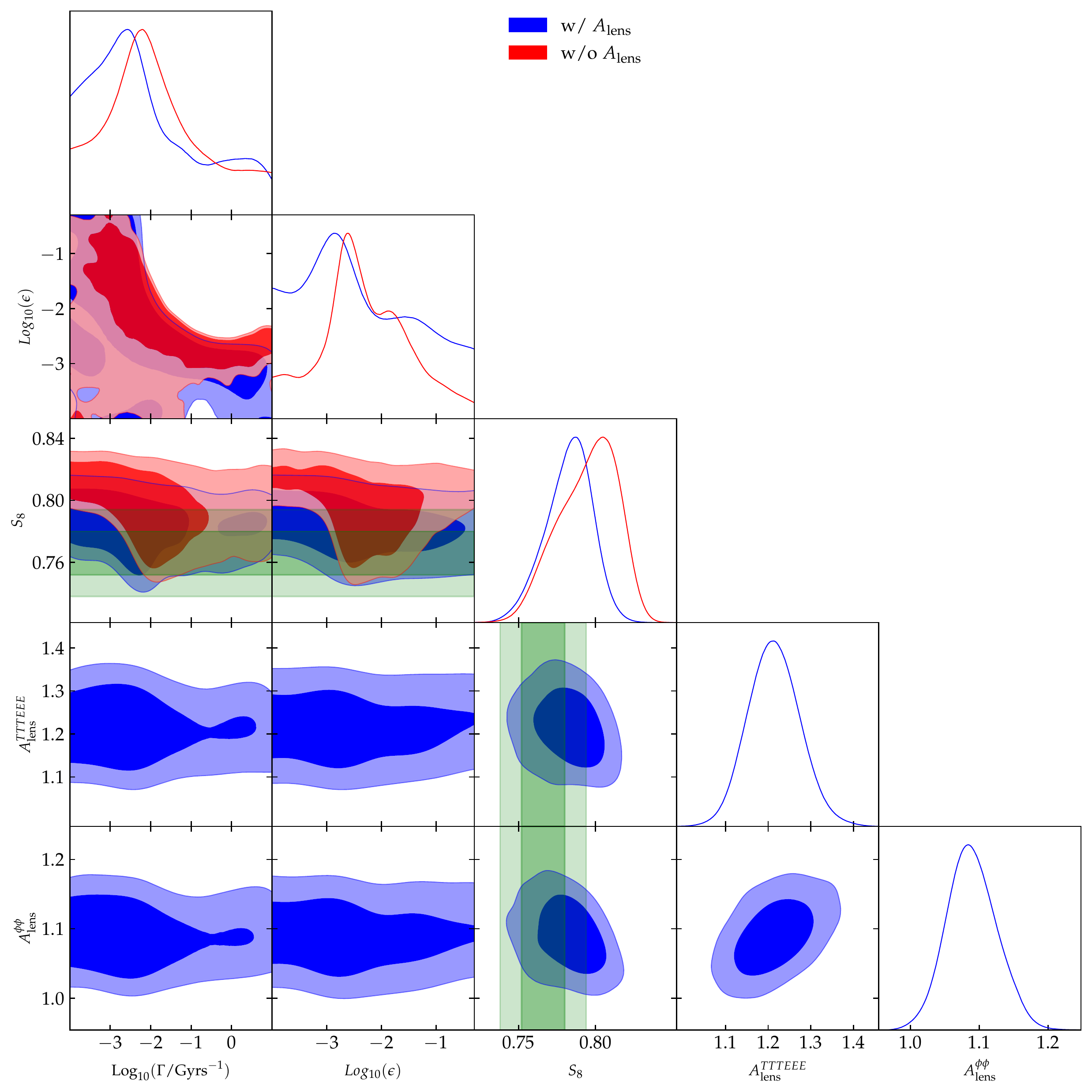}
    \caption{2D posterior distribution of a subset of parameters from our BAO + SNIa + \emph{Planck} + $S_8$ (from KiDS+BOSS+2dFLens) analysis, with and without including the extra $A_{\rm lens}^{\rm TTTEEE}$ and $A_{\rm lens}^{\phi\phi}$ to marginalize over CMB lensing information.}
    \label{fig:alens}
\end{figure}

Another well studied `curiosity' in the recent literature consists in the anomalous amount of lensing estimated from the smoothing of the acoustic peaks at high-$\ell$'s within {\emph{Planck}} data, as quantified by the `$A_{\rm lens}$' parameter \cite{Calabrese:2008rt,Aghanim:2016sns,Aghanim:2018eyx,Efstathiou:2019mdh}. However, this anomalous `lensing' is not supported by the lensing power spectrum reconstruction, such that it is now commonly admitted that this tension (oscillating between the $2-3\sigma$ statistical level) cannot originate from a true lensing effect. On the other hand, it has been understood that this anomaly can be easily resolved if the universe is closed \cite{Handley:2019tkm,DiValentino:2019qzk,Efstathiou:2020wem}, in certain modified gravity theories~\cite{Moshafi:2020rkq},  or in early-universe scenarios inducing a pattern of primordial oscillatory features \citep{Chen:2010xka,Chluba:2015bqa,Slosar:2019gvt,Domenech:2020qay}. Here instead, we wish to check whether this anomaly could impact constraints on the $\Lambda$DDM model, and conversely if the $\Lambda$DDM model could help explaining the existence of such anomalies.  
In fact, it has already been noted that this anomaly could be related to the $S_8$ tension: indeed, once including $A_{\rm lens}$ as an extra free-parameter in the analysis,  it has been shown that the reconstructed cosmology has a smaller $A_s$ and $\omega_{\rm cdm}$ (as well as a higher $H_0$), showing no $S_8$ tension, but a remnant $\sim3.5\sigma$ Hubble tension \cite{Motloch:2018pjy,DiValentino:2018gcu,Motloch:2019gux}.  
To do so, we follow the approach of the SPTpol collaboration and implement two new parameters in {\sf CLASS} that allows to (roughly) marginalize over the lensing information in {\emph{Planck}}. 
The parameter $A_{\rm lens}^{\rm TTTEEE}$ re-scales the amplitude of the lensing power spectrum entering in the high-$\ell$ part of the CMB TT,TE,EE spectra, while the parameter $A_{\rm lens}^{\phi\phi}$ re-scales the amplitude of the lensing power spectrum reconstruction. 
We present the result of a MCMC analysis including {\emph{Planck}} high-$\ell$ TT,TE,EE+lensing+BAO+SNIa+$S_8$ data -- with the two extra lensing parameters -- in Tab.~\ref{tab:alens} and in Fig.~\ref{fig:alens}.
One can see that the  $S_8$ parameter reconstructed once marginalizing over the `$A_{\rm lens}$' anomaly is lower by $\sim0.5\sigma$ than in the baseline analysis. However, the preference for $\Lambda$DDM decreases, with $\text{log}_{10} (\varepsilon)$ now unconstrained. Note also that the $A_{\rm lens}^{\rm TTTEEE}$ is still more than 2$\sigma$ away than the fiducial value $A_{\rm lens}^{\rm TTTEEE} =1$. We can therefore conclude that the $\Lambda$DDM model cannot explain this anomaly and that the preference for $\Lambda$DDM would likely disappear if the $S_8$ tension turns out to be explained by a systematic in {\emph{Planck}} data leading to the anomalous value of the $A_{\rm lens}$ parameters.

\subsubsection{Connections with the recent Xenon1T anomaly}

\begin{figure}
    \centering
    \includegraphics[scale=0.55]{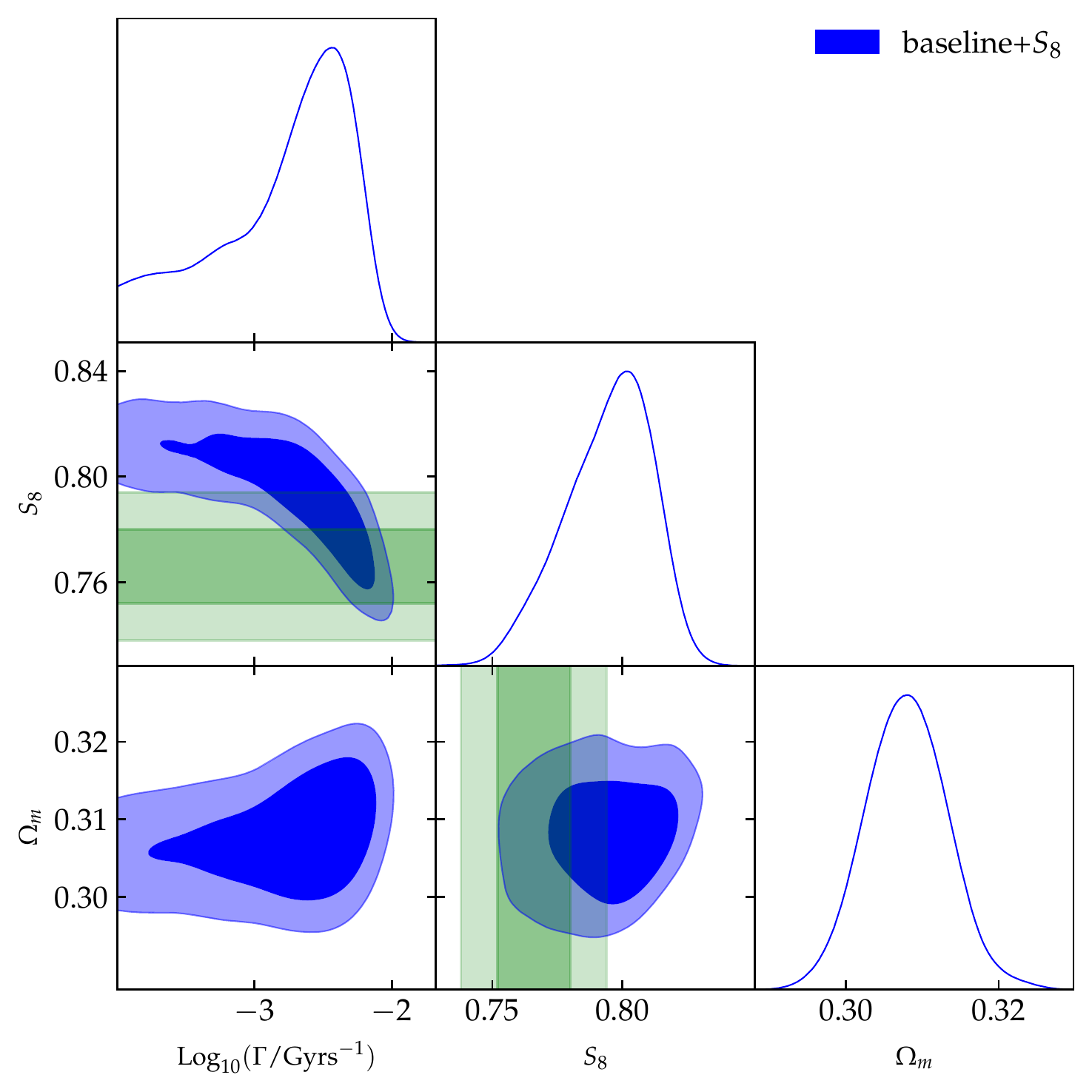}
    \caption{2D posterior distribution of a subset of parameters in our BAO + SNIa + \emph{Planck} + $S_8$ (from KiDS+BOSS+2dFLens) analysis when fixing the DR energy fraction to a value motivated by the recent Xenon1T anomaly ($\varepsilon=0.05$).}
    \label{fig:xenon1t}
\end{figure}

Following our work \cite{Abellan:2020pmw}, let us further study the implications of the $\Lambda$DDM model for the excess of events in the electronic recoils recently reported by the Xenon1T Collaboration \cite{Xenon1tEtAl2020}. It has been pointed out that this excess could {potentially} be explained by the elastic interactions of electrons with a fast DM component of mass $ m \gtrsim 0.1 \ \rm{MeV}$ and velocities $0.05 \lesssim v/c \lesssim 0.13$ \cite{KannikeEtAl2020,Choi:2020udy,Xu:2020qsy} (see also \cite{Buch:2020xyt} for an alternative decaying scenario). Interestingly, the WDM daughter species in the $\Lambda$DDM scenario could {in principle} play the role of such a fast component, since our results indicate that the 1-$\sigma$ range for the speed of the daughter particle  extends up to  $v/c \simeq \varepsilon \simeq 0.05$. In order to test this hypothesis further, we perform another MCMC analysis including {\emph{Planck}} high-$\ell$ TT,TE,EE+lensing+BAO+SNIa+$S_8$, with the DR energy fraction now fixed to $\varepsilon=0.05$. This serves as a proxy for taking into account Xenon1T measurement (alternatively, one could enforce $\varepsilon > 0.05$). The results are summarized in Fig. \ref{fig:xenon1t} and Tab. \ref{tab:xenon1t}. We find best-fit values $\mathrm{Log}_{10} (\Gamma / \mathrm{Gyrs}^{-1}) \simeq -2.4$ and $S_8 \simeq 0.784$, at the cost of a {mild} degradation in the fit to {\emph{Planck}} data ($\Delta\chi^2\simeq+1.7$), indicating that the 2-body decaying scenario {has indeed the potential to provide} a common resolution to the $S_8$ and Xenon1T anomalies.  We leave the construction of a realistic model and study of the subsequent cosmological implications beyond the effect the decay to another study (see \cite{Choi:2020udy,Xu:2020qsy} for examples). 

\begin{table}
 \begin{tabular}{|l|c|} 
 \hline
 Parameter & $\Lambda$DDM ($\varepsilon=0.05$)  \\ 
 \hline
$100 \ \omega_b $ & $ 2.244(2.245)_{-0.015}^{+0.014}$  
\\
$\Odcdm$ & $0.2597(0.2595)_{-0.0067}^{+0.0054}$  
\\
$ H_0 /[{\rm km/s/Mpc}]$ & $67.76(67.78)_{-0.45}^{+0.54}$ \\
$\text{ln}(10^{10} A_s)$ & $3.050(3.047)_{-0.016}^{+0.015}$
\\
$n_s$ & $0.9674(0.9674)_{-0.0039}^{+0.0042}$  
\\
$\tau_{\rm reio}$ & $0.0573(0.0559)_{-0.0079}^{+0.0073}$  \\
$\text{log}_{10} (\Gamma/[{\rm Gyr}^{-1}])$	& $-2.72(-2.44)_{-0.21}^{+0.61}$ 
\\ 
$\Omega_{\rm m}$ & $0.3093(0.3090)_{-0.007}^{+0.0057}$ \\
$S_8$	& $0.794(0.786)_{-0.015}^{+0.021}$  \\
\hline
$\chi^2_{\rm min}$ & 2057.6   \\ 
 \hline
\end{tabular}

\caption{The mean (best-fit) $\pm 1\sigma$ errors of the cosmological parameters from our {\emph{Planck}} high-$\ell$ TT,TE,EE+lensing+BAO+SNIa+$S_8$ analysis, when fixing the DR energy fraction to $\varepsilon=0.05$. We also report the best-fit $\chi^2$.}
\label{tab:xenon1t}
\end{table}

\section{\label{sec:cmbs4}Detecting DDM in the CMB: impact of current and future data}
{In this section, we confront the DDM model to additional CMB data from current ground based surveys and perform forecast for future surveys. In addition to Planck, we consider first, the high-$\ell$ CMB EE and TE ($50 \leq \ell \leq 8000$)~\cite{Henning:2017nuy} measurements and the reconstructed gravitational lensing potential ($100 \leq \ell \leq 8000$)~\cite{Bianchini:2019vxp} from the 500deg SPTpol survey~\cite{Chudaykin:2020acu}.  We then include the high-$\ell$ CMB TT, EE and TE ($ 350 \leq \ell \leq 4125$) data from the DR4 of the ACTPol survey~\cite{Aiola:2020azj,Choi:2020ccd}. Finally, we demonstrate that an experiment like CMB-S4 can unambiguously detect the DDM model.}
\subsubsection{Confronting $\Lambda${\rm{DDM}} to SPTpol data}
\begin{figure}
    \centering
    \includegraphics[scale=0.42]{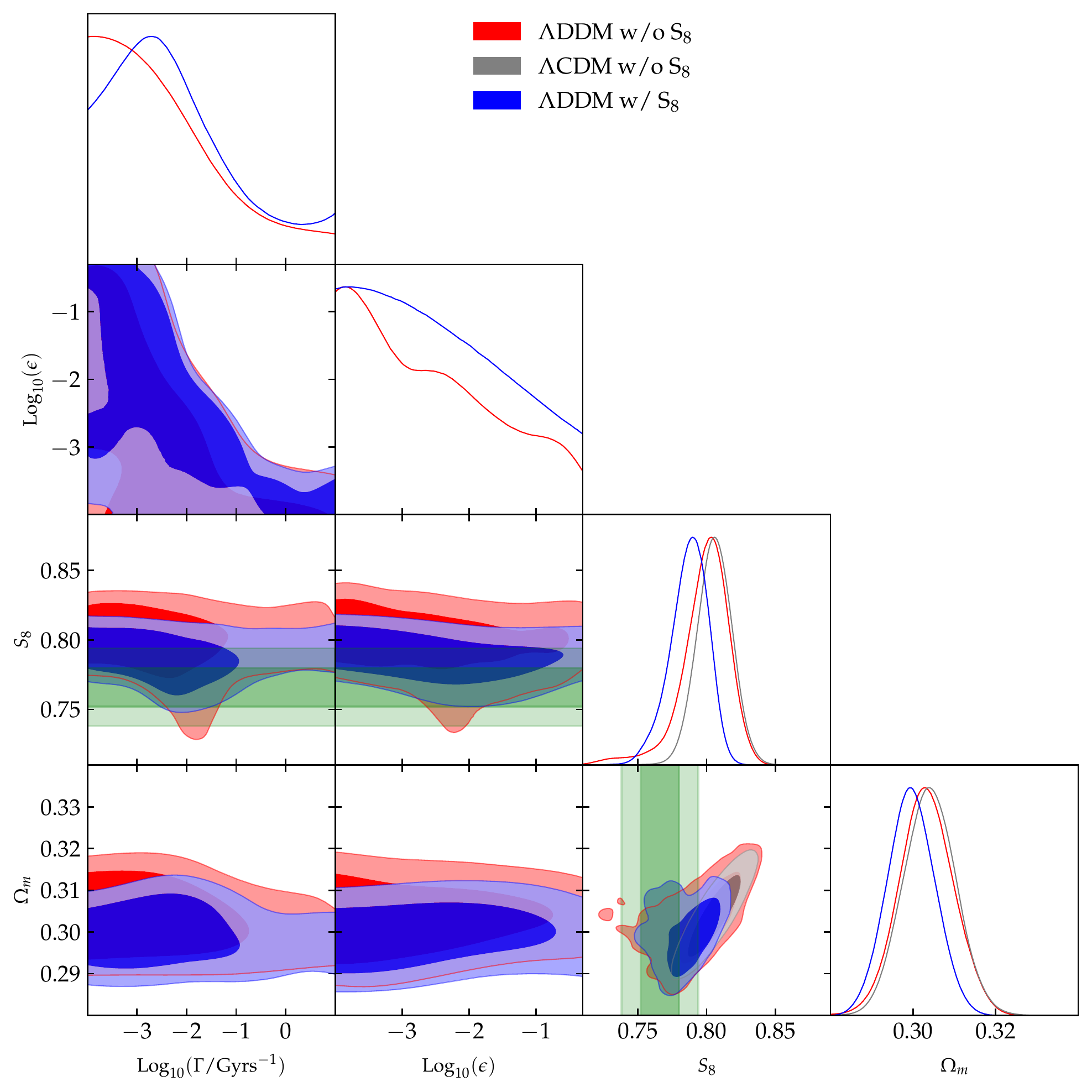}
    \caption{2D posterior distribution of a subset of parameters in the joint BAO + SNIa + {\emph{Planck}} + SPTpol analysis, with and without imposing a prior on S$_8$ from KiDS+BOSS+2dFLens, compared to the $\Lambda$CDM scenario.}
    \label{fig:spt1}
\end{figure}

\begin{figure}
    \centering
    \includegraphics[scale=0.42]{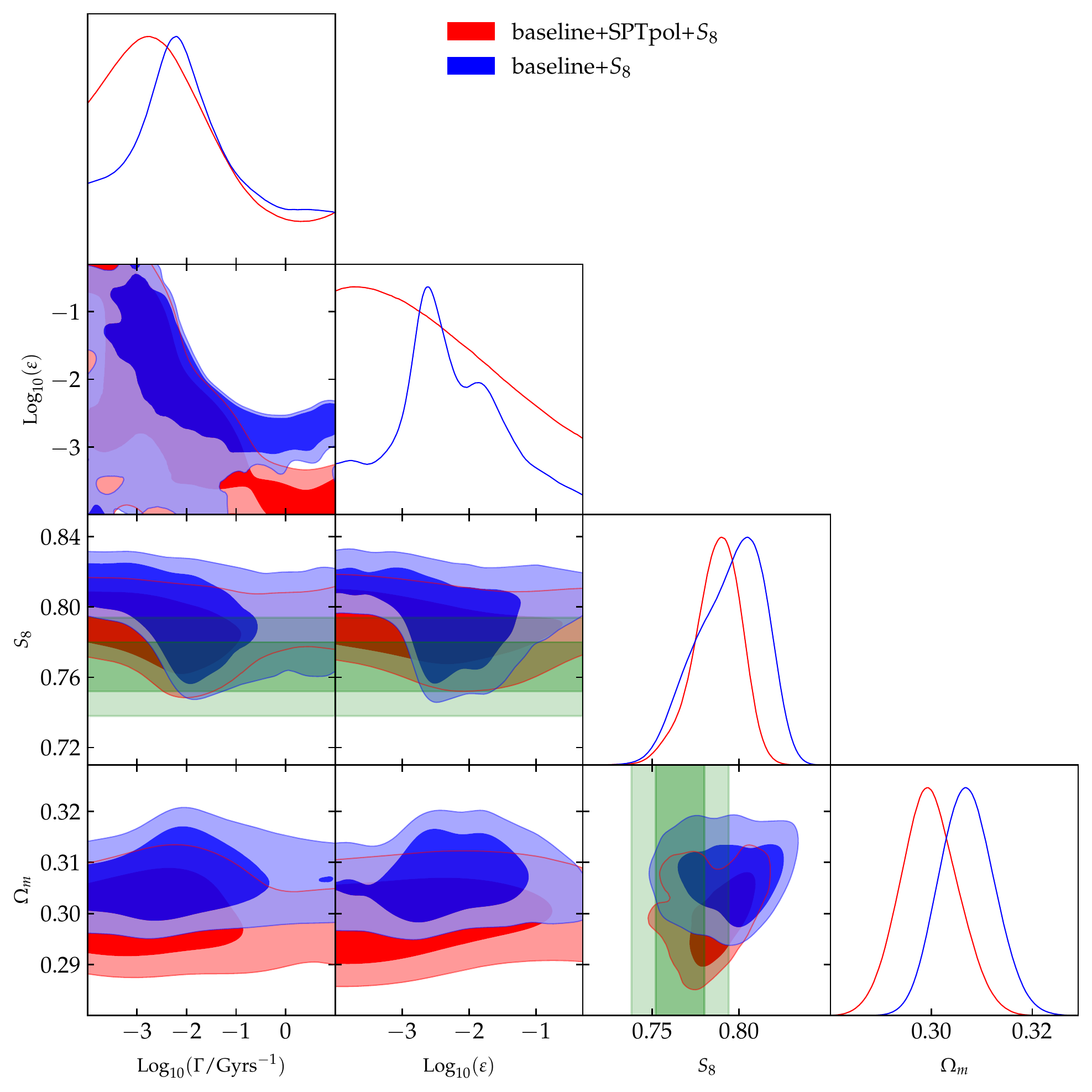}
    \caption{2D posterior distribution of a subset of parameters in our from our BAO + SNIa + \emph{Planck} + $S_8$ (from KiDS+BOSS+2dFLens) analysis, compared to the joint BAO + SNIa + {\emph{Planck}} + SPTpol + $S_8$ analysis.}
    \label{fig:spt2}
\end{figure}

\begin{table}
 \scalebox{0.95}{
 \begin{tabular}{|l|c|c|} 
  \hline

 Data &  \multicolumn{2}{c|}{BAO + SNIa + \emph{Planck}TT + SPTpol}\\ 

 \hline
 Parameter & w/o $S_8$ & w/ $S_8$ \\ 
 \hline
$100 \ \omega_b $								& $2.239(2.2378)_{-0.018}^{+0.019}$ & $2.241(2.247)_{-0.019}^{+0.016}$ \\
$\Omega_{\rm dcdm}^{\rm ini}$ 					& $0.2544(0.2557)_{-0.0061}^{+0.0057}$ & $0.2514(0.2532)_{-0.0055}^{+0.0054}$ \\
$ H_0 /[{\rm km/s/Mpc}]$ 						& $68.15(68.03)_{-0.48}^{+0.51}$ & $68.39(68.25)_{-0.46}^{+0.47}$ \\
$\text{ln}(10^{10} A_s)$						& $3.031(3.026)_{-0.017}^{+0.016}$ & $3.026(3.018)_{-0.014}^{+0.017}$ \\
$n_s$											& $0.9701(0.9695)_{-0.0042}^{+0.0041}$ & $0.9712(0.9707)_{-0.004}^{+0.0037}$ \\
$\tau_{\rm reio}$								& $0.0509(0.0481)_{-0.0081}^{+0.0074}$ & $0.0494(0.0457)_{-0.0073}^{+0.0082}$ \\
$\text{log}_{10} (\Gamma/[{\rm Gyr}^{-1}])$		& $-2.38(-1.73)_{-1.6}^{+0.38}$ & $-2.25(-1.35)_{-1.7}^{+0.42}$ \\
$\text{log}_{10} (\varepsilon)$					& unconstrained(-2.73)& unconstrained(-2.57)\\
$\Omega_{\rm m}$								& $0.3033(0.3046)_{-0.0067}^{+0.0064}$ & $0.2999(0.3013)_{-0.0062}^{+0.0059}$ \\
$S_8$										& $0.799(0.798)_{-0.015}^{+0.022}$  &  $0.787(0.767)_{-0.013}^{+0.016}$\\
\hline
 $\chi^2_{\rm min}$ & 1816.3 & 1816.8  \\
 \hline
\end{tabular} }
\caption{The mean (best-fit) $\pm 1\sigma$ errors of the cosmological parameters from our BAO + SNIa + \emph{Planck}TT + SPTpol analysis, with and without imposing a split-normal likelihood on $S_8$ (from KiDS+BOSS+2dFLens). For each data-set, we also report the best-fit $\chi^2$.}
\label{tab:spt}
\end{table}

It is interesting to test the robustness of the DDM ``detection'' to a change of CMB data sets, especially given the impact of marginalizing over the `$A_{\rm lens}$' anomaly as discussed in previous section. We thus confront the $\Lambda$DDM scenario under study against a set of CMB data constituted by low-$\ell$ temperature and polarization as well as high-$\ell$ temperature data from {\emph{Planck}}, in combination with high-$\ell$ polarization data from SPTpol (see section~\ref{sec_dcdm:data} for further details and references). It has been shown indeed that such a joint analysis predicts an amount of CMB lensing consistent with the $\Lambda$CDM expectation~\cite{Chudaykin:2020acu},~i.e.~no `$A_{\rm lens}$' anomaly, and no $S_8$ tension.  This is manifest in Fig.~\ref{fig:spt1}, where we compare predictions from the $\Lambda$CDM and the $\Lambda$DDM models, the latter both with and without including information on $S_8$ from KIDS1000+BOSS+2dFLens.
As one can easily see, both cosmological models predict a $S_8$ value in excellent agreement with the KIDS1000+BOSS+2dFLens measurement, displayed as a green horizontal band. Our results regarding the $\Lambda$DDM model are also reported in Tab.~\ref{tab:spt}.

In Fig.~\ref{fig:spt2}, instead, we report a comparison between our baseline $\Lambda$DDM analysis and the {\emph{Planck}}+SPT one. First and foremost, SPTpol appears in very good agreement with the $\Lambda$DDM model resolution of the $S_8$ tension required by {\em Planck}. However, the $\Lambda$DDM parameters are largely unconstrained in that case and no deviations from $\Lambda$CDM are visible, which further establishes that if the $S_8$ tension turns out to be explained by a systematic in {\emph{Planck}} high-$\ell$ polarization data, the preference for $\Lambda$DDM is likely to vanish.

\begin{figure}
    \centering

    \includegraphics[scale=0.45]{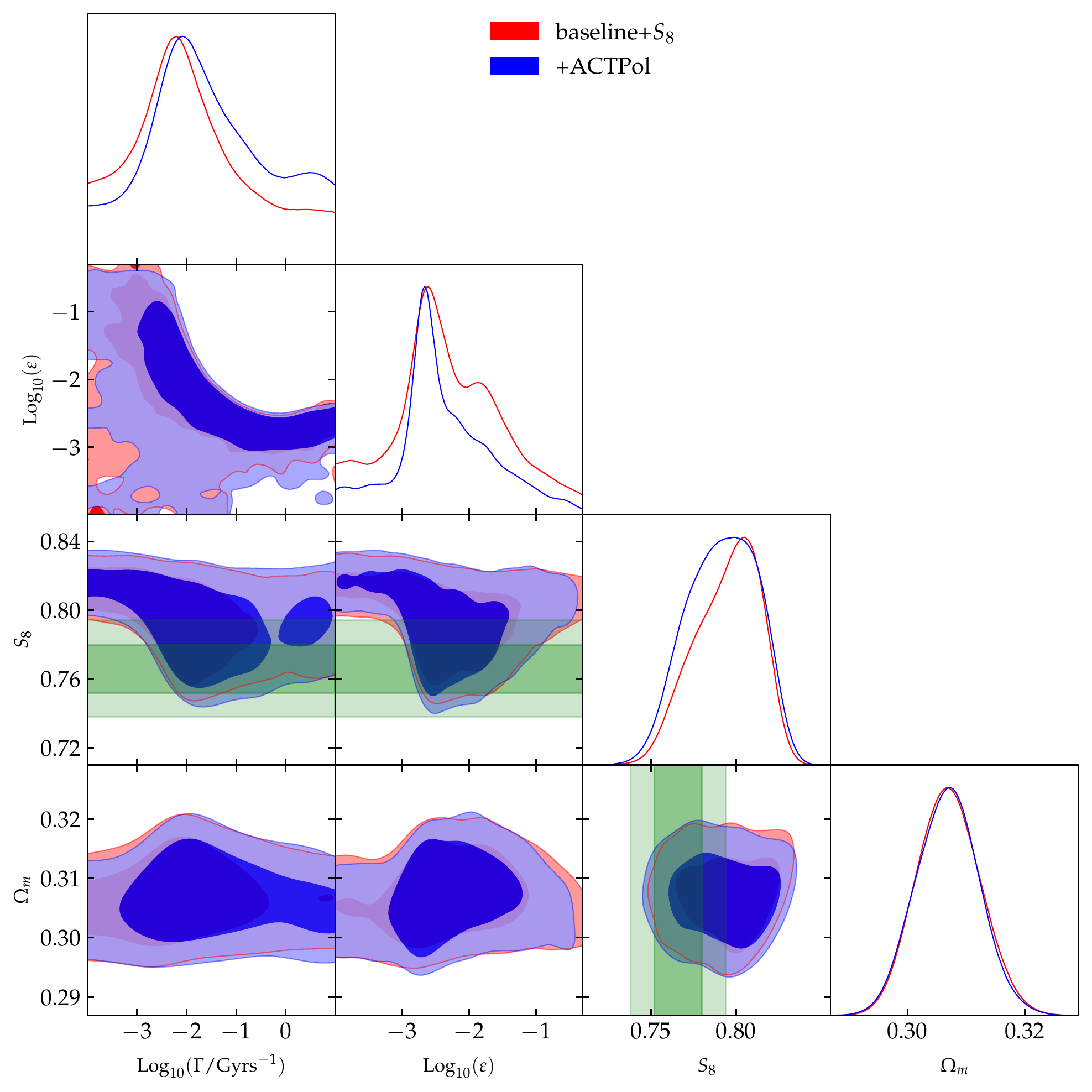}
    \caption{2D posterior distribution of a subset of parameters in the case BAO + SNIa + \emph{Planck} + $S_8$ (from KiDS+BOSS+2dFLens), with and without the inclusion of ACT data.}
    \label{fig:ACT-vs-Planck}
\end{figure}

{\subsubsection{Confronting $\Lambda${\rm{DDM}} to ACTPol data}}
{{
Second, we confront the $\Lambda$DDM model to the combination of Planck and ACTPol data, to test whether more accurate measurements at high-$\ell$ can further constrain the model. 
Within $\Lambda$CDM, it has been found that ACTPol data (when combined with WMAP) also favor relatively high $S_8$, in $2.1\sigma$ tension with KiDS1000+BOSS+2dFLens \cite{Aiola:2020azj}. To limit double counting of information, we follow the procedure of the ACT collaboration and truncate multipoles $\ell < 1800$ in the ACT TT data. The results of this analysis are presented in Fig. \ref{fig:ACT-vs-Planck}. and Table \ref{tab:act}. Interestingly, the $\Lambda$DDM parameter $\varepsilon$ is more precisely measured with the inclusion of the ACTPol data, while the mean value is barely affected. Compared with $\Lambda$CDM, the $\Delta\chi^2$ in favor of $\Lambda$DDM is now  $-6.7$, and the level of tension between Planck+ACT+BAO+SN1a and $S_8$ from KIDS1000+BOSS+2dFLens is $1.3\sigma$.
We conclude that ACT data are in very good agreement with the $\Lambda$DDM model, slightly increasing its preference over $\Lambda$CDM.}}

\begin{table}
 \scalebox{0.95}{
 \begin{tabular}{|l|c|c|} 
  \hline

 Data &  \multicolumn{2}{c|}{BAO + SNIa + \emph{Planck} + ACTPol}\\ 

 \hline
 Parameter & w/o $S_8$ & w/ $S_8$ \\ 
 \hline
$100 \ \omega_b $			&	$2.245( 2.242)\pm0.013$	& $2.243(2.243)\pm0.012$ \\
$\Omega_{\rm dcdm}^{\rm ini}$ &	$0.26(0.261)\pm0.005$	& $0.2583(0.2591)_{-0.0051}^{+0.0048}$\\
$ H_0 /[{\rm km/s/Mpc}]$ 	&$67.77(67.73)_{-0.42}^{+0.33}$		& $67.92(67.84)\pm0.40$ \\
$\text{ln}(10^{10} A_s)$&$3.058(3.054)\pm0.015$	& $3.056(3.061)\pm0.015$ \\
$n_s$			&	$0.9703(0.9697)_{-0.004}^{+0.0034}$	& $0.9708( 0.9706)\pm0.0036$\\
$\tau_{\rm reio}$		& $0.0571(0.0549)_{-0.0084}^{+0.0069}$	& $0.0567(0.0592)_{-0.0075}^{+0.007}$ \\
$\text{log}_{10} (\Gamma/[{\rm Gyr}^{-1}])$&$0.92(-2.92)_{-1.90}^{+0.55}$	&  $-1.56(-1.19)_{-1.50}^{+1.10}$\\
$\text{log}_{10} (\varepsilon)$ &unconstrained (-3.88)& $-2.34(-2.54)\pm0.65$\\
$\Omega_{\rm m}$	&	$0.3094(0.3108)_{-0.0054}^{+0.0051}$	&  $0.3069(0.3068)_{-0.0056}^{+0.0050}$ \\
$S_8$			&	$0.822(0.829)_{-0.011}^{+0.018}$			&$0.792(0.772)_{-0.019}^{+0.025}$  \\
\hline
 $\chi^2_{\rm min}$ & 2294.54& 2296.20\\
 \hline
$\Delta\chi^2(\Lambda\text{CDM})$ & -0.12 & -6.7\\
 \hline
\end{tabular} }
\caption{The mean (best-fit) $\pm 1\sigma$ errors of the cosmological parameters from our BAO + SNIa + \emph{Planck} + ACTPol analysis, with and without imposing a split-normal likelihood on $S_8$ (from KiDS+BOSS+2dFLens). For each data-set, we also report the best-fit $\chi^2$ and the $\Delta\chi^2(\Lambda\text{CDM})\equiv \chi^2_{\rm min} (\Lambda\text{DDM}) -\chi^2_{\rm min} (\Lambda\text{CDM})$.}
\label{tab:act}
\end{table}

\vspace{5mm}

{\subsubsection{{Towards detecting the $\Lambda$DDM model with CMB-S4}}}

\begin{figure}
    \centering
    \includegraphics[scale=0.44]{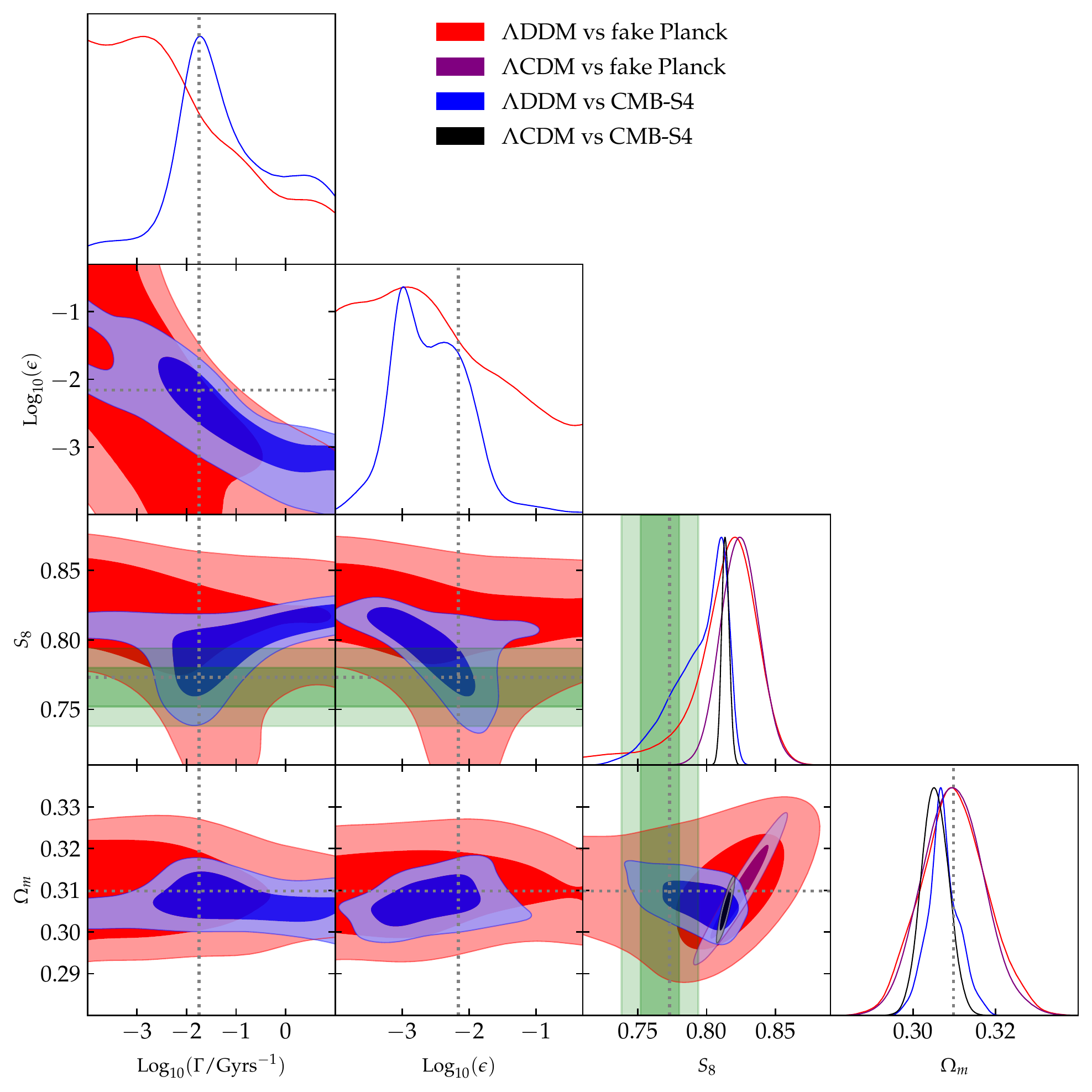}
    \caption{2D posterior distribution of a subset of parameters reconstructed from a fit to simulated
    $Planck$ and CMB-S4 data. The fiducial model has ${\rm log}_{10}(\epsilon) = -2.16$ and ${\rm log}_{10}(\Gamma/[{\rm Gyr^{-1}}]) = -1.74$, as denoted by the gray dotted lines.}
    \label{fig:fakeplanck}
\end{figure}

{{
As we have extensively discussed, current CMB data are not sensitive enough to detect DDM, so that the preference for non-standard values for $\varepsilon$ and $\Gamma$ is fully driven by the inclusion of the $S_8$ measurement from weak lensing data in the analysis. To further stress this aspect we have performed additional analyses fitting a set of mock CMB data generated starting from our reference best-fit $\Lambda$DDM model (i.e.,~that from the $Planck$+BAO+SNIa+KiDS+BOSS+2dFLens run, reported in the first column of Tab.~\ref{tab:s8}).
The resulting contour plots are shown in Fig.~\ref{fig:fakeplanck}, where we compare the constraints that $Planck$ would obtain if the ``true'' cosmological model actually contained DDM, with those that a future generation CMB survey (CMB-S4) would get. Concretely, this task was pursued by using the `Planck-fake-realistic' and `CMB-S4' likelihoods available in \texttt{MontePython-v3}. The former allowed us to generate synthetic Planck data, whereas the latter includes multipoles $\ell$ from 30 to 3000, assuming a sky coverage of 40$\%$, uncorrelated Gaussian error on each $a_{\ell m}$’s, uncorrelated temperature and polarization noise, and perfect foreground cleaning up to $\ell_{\rm max}$\footnote{To overcome the lack of low-$\ell$ data in the CMB-S4 analysis, we have imposed a Gaussian prior on the optical depth to reionization, centred on its best-fit value from our reference $Planck$+BAO+SNIa+KiDS+BOSS+2dFLens analysis, namely $\tau_{\rm reio} = 0.0582 \pm 0.008$.}.  All details about the likelihood can be found in Tab.~1 of Ref.~\cite{Brinckmann:2018owf}.

As expected, from Fig.~\ref{fig:fakeplanck} it is manifest that $Planck$ alone could not detect DDM even if its signature was truly imprinted in CMB data. Note indeed that the red contours barely features an overlap at 2$\sigma$ in $S_8$, and only upper limits on ${\rm Log}_{10}\Gamma$, ${\rm Log}_{10}\varepsilon$. This, as we explained earlier, is a consequence of the degeneracy that exists within $\Lambda$CDM and leads to a bias in the Bayesian analysis.
Therefore, the information that matters in quantifying the success of the resolution is rather contained in the $\chi^2$ values:
just like in the analysis of real data, we find that, when testing the $\Lambda$CDM model against the mock data that contains the $\Lambda$DDM signal, the $\chi^2$
in the $\Lambda$CDM model is identical to that of the $\Lambda$DDM model. This clearly shows that Planck cannot disentangle between $\Lambda$CDM and $\Lambda$DDM, while the $S_8$ measurements favors $\Lambda$DDM (in terms of $\chi^2$).
This is not the case for CMB-S4, which contours are over-plotted in blue: if the real Universe contains DDM, CMB-S4 would unequivocally detect its signature, finding $\varepsilon \neq 0$ at $\gtrsim$ 2 $\sigma$ level. The $\Delta \chi^2$ in favour of the $\Lambda$DDM model {\em from (mock) CMB-S4 data alone} is in fact $\simeq +8$}}.

\vspace{5mm}

\section{\label{sec_dcdm:conclusion}Conclusions}

In this paper, we have performed a comprehensive cosmological study of the CDM 2-body decay scenario dubbed `$\Lambda$DDM', whereby decays are characterized both by the decay rate $\Gamma$ and energy fraction converted to radiation $\varepsilon$, including 
a fully consistent treatment of the linear perturbations of the WDM daughter component.

To that end, {we have made use of a new approximation scheme, introduced in Ref.~\cite{Abellan:2020pmw}}, that allows to accurately and quickly compute the dynamics of the WDM linear perturbations by treating the WDM species as a viscous fluid. {Close to the best-fit values}, our approximation scheme is accurate at the ${\cal O}(0.1\%)$ level in the CMB power spectra and ${\cal O}(1\%)$ level in the linear matter power spectrum (see App. \ref{sec_dcdm:appendix_numerics}). 

We have then discussed in detail the dynamics of linear density perturbations of the mother and daughter particles, as well as the physical effects of the $\Lambda$DDM model on the CMB and matter power spectra.
We have shown that accurate CMB (lensing in particular) and matter power spectrum measurements have the potential to detect both the decay rate $\Gamma$ and energy fraction converted to radiation $\varepsilon$.

In a second part, we have performed a set of MCMC analyses of the $\Lambda$DDM model against a suite of up-to-date low- and high-redshift data-sets. 
We have compared the constraints obtained from BAO and SNIa data, thereby solely based on background effects, to those obtained from the full {\em Planck} data-set, that requires instead an accurate description of the WDM linear perturbations. We find that {\emph{Planck}} CMB data constrain the $\Lambda$DDM model $\sim 1$ order of magnitude better than current BAO+SNIa data. However, we also show that despite these constraints, the $\Lambda$DDM model provides a promising possibility to resolve the $S_8$ tension, as detailed in a previous paper~\cite{Abellan:2020pmw}.

We have then tested the robustness of the $\Lambda$DDM resolution to the $S_8$ tension to a number of change in the analysis. 
We show that the mild preference for the $\Lambda$DDM model over $\Lambda$CDM is tied to the $S_8$ value chosen in the analysis.
Concretely, the $S_8$ value from the KiDS+Viking+DES analysis, which has a higher level of tension with the {\emph{Planck}} $\Lambda$CDM prediction than the baseline KiDS+BOSS+2dFLens value, leads to a stronger preference in favor of  the $\Lambda$DDM model. However, the DES-only result, which is in reasonable agreement with {\emph{Planck}}, leads to a weaker preference of the $\Lambda$DDM model. 
Similarly, once marginalizing over the lensing information in {\emph{Planck}} through the $A_{\rm lens}$ parameter (we used two extra parameters in practice describing the normalization of the lensing power spectrum and the normalization of the lensing smoothing effect in the high-$\ell$ TT,TE,EE power spectra), or when trading the {\emph{Planck}} high-$\ell$ TE,EE power spectra for the SPTpol ones, the preference for the $\Lambda$DDM model decreases.
This is because in these two cases the inferred $\Lambda$CDM model has a smaller $S_8$ value, showing less of a tension with the weak lensing surveys. 
This indicates that if the $S_8$ tension ends up coming from an unknown systematic within weak lensing surveys or within {\emph{Planck}} data, the preference for the $\Lambda$DDM model would likely disappear. {On the other hand, when combining {\emph{Planck}} with ACTPol the mild preference for $\Lambda$DDM increases, and the remaining `tension' with $S_8$ is now only $\sim1.3\sigma$. }

We have also tested the intriguing possibility that the recent Xenon1T excess is due to the $\Lambda$DDM model.
To that end, we have performed a additional MCMC analysis fixing $\eps=0.05$ as required by Xenon1T. 
We find that it is easy to resolve the $S_8$ tension in that case, pointing to a DCDM lifetime of $\text{log}_{10} (\Gamma/[{\rm Gyr}^{-1}])=-2.72_{-0.21}^{+0.61}$. Interestingly, this comes at the cost of a very minor degradation in {\em Planck} fit ($\Delta\chi^2\simeq+1.7$), indicating that {\em Planck}, BAO and SNIa data are in good agreement with this model.

{Finally, by performing an analysis where we artificially introduce a DDM signal in a set of mock CMB data, we explicitly demonstrate that while current CMB data alone are not sensitive enough to distinguish between standard CDM and  DDM, next-generation CMB experiments (CMB-S4) can unambiguously detect its signature.}

It will be very interesting to go beyond the linear aspects discussed in this work and study the non-linear evolution of density perturbations, in order to be able to make use of the full power of the KiDS, BOSS and DES likelihoods. This could for instance be done with N-body simulations, as in Refs.~\cite{Wang:2012eka,Wang:2013rha,Wang:2014ina}, or via the Effective Theory of LSS \cite{DAmico:2020kxu,Chudaykin:2020aoj}. This will be even more crucial with upcoming surveys such as  Euclid \cite{Amendola:2016saw}, LSST \cite{Mandelbaum:2018ouv}, and DESI \cite{Aghamousa:2016zmz}, which will measure the matter power spectrum and the growth factor with great accuracy up to $z\sim2$. It might also be possible to test the $\Lambda$DDM model with current Lyman-$\alpha$ forest flux power spectrum data  \cite{Wang:2013rha,Murgia:2017lwo,Murgia:2018now,Archidiacono:2019wdp,Miller:2019pss,Enzi:2020ieg}.  We plan to study non-linear aspects and the discovery potential of these surveys in an upcoming work.

\begin{acknowledgements}

The authors are thankful to Julien Lavalle and Rodrigo Calder\'on for many useful comments and discussions. The authors acknowledge the use of computational resources from the CNRS/IN2P3 Computing Centre (CC-IN2P3) in Lyon, the IN2P3/CNRS and the Dark Energy computing Center funded by the OCEVU Labex (ANR-11-LABX-0060) and the Excellence Initiative of Aix-Marseille University (A*MIDEX) of the “Investissements d’Avenir” programme. This project has received support from the European Union’s Horizon 2020 research and innovation program under the Marie Skodowska-Curie grant agreement No 860881-HIDDeN.

\end{acknowledgements}

\appendix

\section{\label{sec_dcdm:appendix_numerics}Numerical implementation and accuracy of the WDM fluid approximation}

\begin{figure}
\includegraphics[scale=0.3]{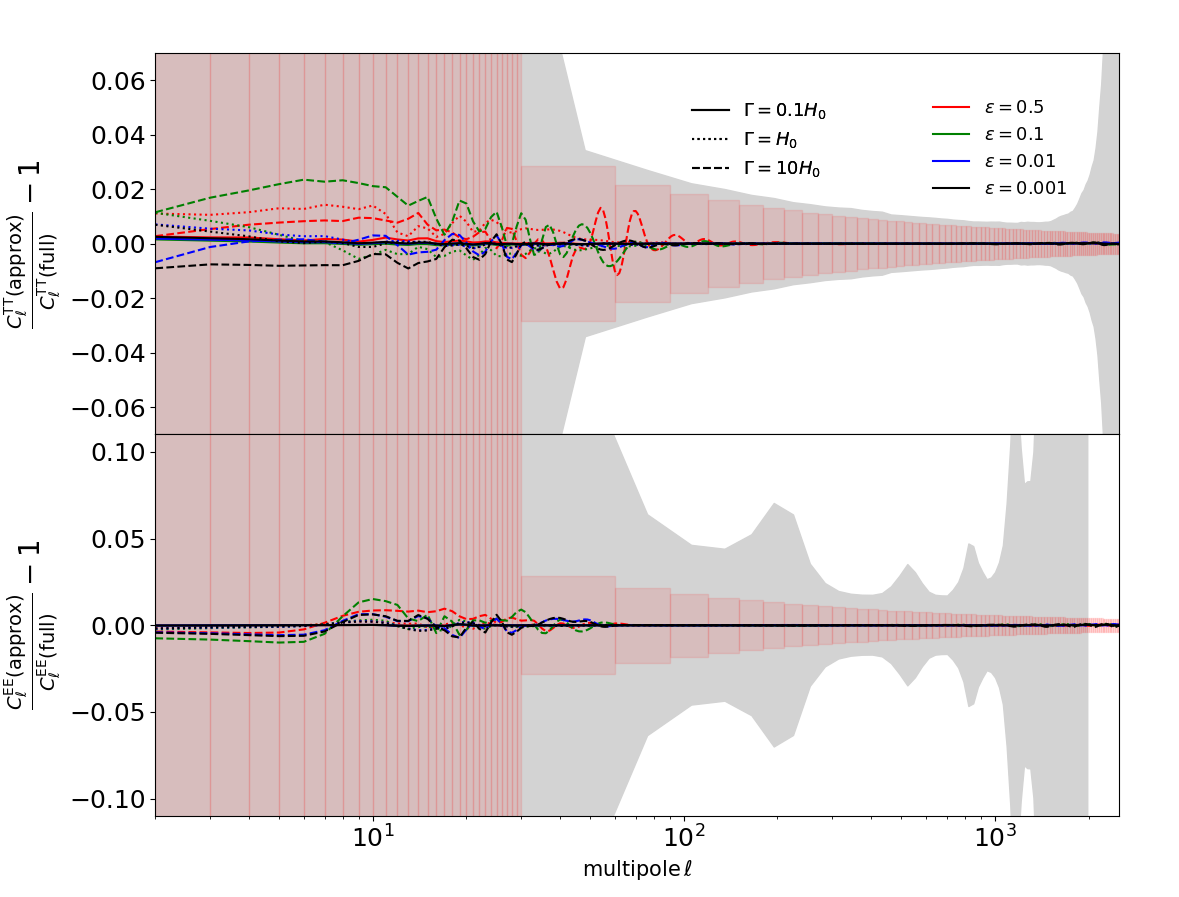}
\caption{\label{fig:fluid_residuals_1} Residuals of the lensed CMB TT power spectrum (upper) and EE power spectrum (lower) in the WDM fluid approximation, with respect to the full hierarchy calculation, for a grid of values covering most of the parameter space: $\Gamma/H_0 = 0.1, 1, 10 $ and $\varepsilon=0.5, 0.1, 0.01, 0.001 $.  The gray shaded regions indicate {\emph{Planck}} 1$\sigma$ errors, while the pink shaded areas indicate cosmic variance up to $\ell =3000$.} 
\end{figure}

\begin{figure}
\includegraphics[scale=0.42]{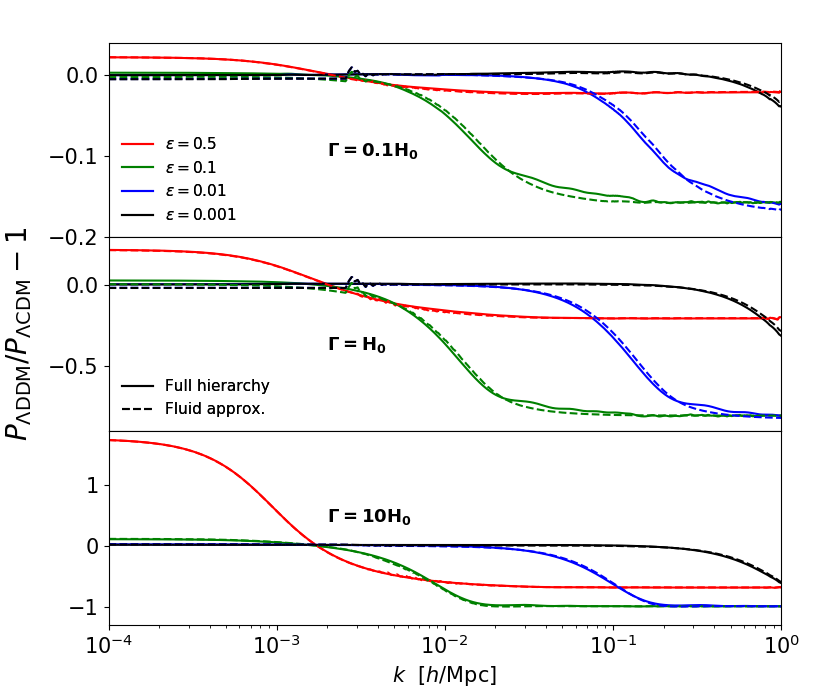}
\caption{\label{fig:fluid_residuals_2} Residuals of the linear matter power spectrum (at $z=0$), with respect to the baseline $\Lambda$CDM model, for the same grid of parameter values considered in Fig. \ref{fig:fluid_residuals_1}, both from the full hierarchy calculation (solid lines) and the WDM fluid approximation(dashed lines).  }
\end{figure}

In order to solve the cosmic evolution equations derived in section \ref{sec_dcdm:model}, we modified the publicly available numerical Boltzmann solver \texttt{CLASS} \cite{Blas_2011, Lesgourgues:2011rh}. We now briefly illustrate our implementation of the $\Lambda$DDM model.

First let us notice that, when solving the background equations for all cosmological species, the DE abundance is iteratively derived through the budget equation, $\Omega_{\Lambda}= 1-\sum_i \Omega_i$, where the sum includes the current abundance of all other components, which are not known a priori. 

We thus applied a \textit{shooting method} for the aforementioned parameter, i.e. we guess an initial $\Omega_\Lambda$, we solve the system of background equations to obtain $\sum_i \Omega_i$, and re-compute $\Omega_\Lambda$. The procedure is iterated until convergence is achieved. 
The WDM density is computed by solving Eq. \eqref{rhowdm} in 2800 momentum bins, approximately as many as the time-steps used to describe its background evolution.

At the linear perturbation level, we truncate the hierarchy of the PSD multipole equations 
for both the daughter particles at a $\ell_{\rm max} =17$. 
We set the initial conditions for the WDM species following the same procedure of \cite{Aoyama:2014tga}. 

On conformal times $\tau < \tau_q$, we set all $\Delta f_{\wdm,\ell} = 0$, since no daughter particle with comoving momentum $q$ could have been produced. 
On the crossing time $\tau=\tau_q$, one should be more careful, as the terms with $\bar{f}_\wdm$ in Eqs. \eqref{delta_f_0} and \eqref{delta_f_2} contain a Dirac delta and, when integrated, a Heaviside function. Thus, the corresponding initial conditions for $\Delta f_{\wdm,0} (\tau_q)$ and $\Delta f_{\wdm,2} (\tau_q)$ are not-vanishing. We set them according to the analytical formulas (A.5) and (A.7) from \cite{Aoyama:2014tga}. 

Finally, on times $\tau > \tau_q$, we treat the WDM component as a massive neutrino species, and we solve the corresponding hierarchy of equations in 300 momentum-bins. This number of bins is chosen simply because it gives a good compromise between speed and accuracy: it is large enough to accurately describe 
the super-Hubble and Hubble-crossing scales, where the dynamics is relatively simple, and small enough to not become too computationally expensive \footnote{
 Note that the number of bins used at both the background and perturbation level is much larger than the one typically used in standard \texttt{CLASS} analysis for massive active neutrinos, given that the time-dependence of the background PSD of the WDM requires a finer momentum resolution. Regarding the momentum spacing, we have considered  a logarithmic Simpson quadrature instead of the Gauss-Laguerre quadrature typically used in standard \texttt{CLASS} analysis.  }.
On sub-Hubble scales, when $k \tau $ is larger than a threshold value $(k \tau)_{\rm fluid}$, we switch-on the fluid approximation described in section \ref{sec_dcdm:perturbation2}. 
The WDM dynamics is now described by Eqs. \eqref{delta_dot_wdm}-\eqref{sound_speed} and \eqref{sync_sound_speed}. We have chosen $(k \tau)_{\rm fluid} = 25$ to provide the speed yet accurate enough for the purposes of the current analyses.

In Figs.~\ref{fig:fluid_residuals_1} and \ref{fig:fluid_residuals_2} we explicitly compare the novel approximation scheme with the results of the ``exact'' computation for the WDM species. For the latter, we solve the full Boltzmann hierarchy using $10^4$ momentum-bins and $\ell_{\rm max}=17$. In Fig. \ref{fig:fluid_residuals_1} we show the residuals of the lensed CMB TT and EE power spectra in the WDM fluid approximation, with respect to the full computation, for a grid of parameter values given by
$\Gamma/H_0 = 0.1, 1, 10 $ and $\varepsilon=0.5, 0.1, 0.01, 0.001 $. 
These values span most of the parameter space in the $\Lambda$DDM framework, and for none of them the residuals exceed the {\emph{Planck}}  $1\sigma$ uncertainties, which are indicated by the gray shaded regions, nor the error bars for a cosmic-variance-limited experiment (close to CMB-S4 errors), indicated by the pink shaded regions.

The predictions for the linear matter power spectrum $P(k)$ are less accurate than for the anisotropy spectra, because the former is more sensitive to the dynamics of the daughter particles.
Close to the best-fit parameter values, and in general close to $\Lambda$CDM, the residual errors between the full and the fluid calculations in $P(k)$ are $\mathcal{O}(1\%)$, but they can become higher far away from the best-fit. In particular, we have verified that inside the parameter region delimited by $\rm{Log}_{10} \varepsilon \in [-2.3, -0.7]$ and $\mathbf{\rm{Log}_{10} (\Gamma /\rm{Gyrs}^{-1}) \in [-1.3, 1]} $, the residual errors are typically larger than $10 \%$, so the fluid approximation should be used with caution in this region. However, this portion of the parameter space is deeply inside the $2\sigma$ exclusion region, as one can check by looking at Fig. \ref{fig:Prior}. In addition, given that current data are mostly sensitive to integrals over $P(k)$ (e.g. $S_8$, CMB lensing), we are mainly interested in getting accurate predictions for the departures from $\Lambda$CDM (rather than the exact shape of the matter spectrum itself). To illustrate that, we have computed the residuals of the linear matter power spectrum (at $z=0$) with respect to our baseline $\Lambda$CDM model, for both the fluid and the full hierarchy calculations. The results are shown in Fig. \ref{fig:fluid_residuals_2}, where we have spanned the same parameter values as in Fig. \ref{fig:fluid_residuals_1}. We can see that, for all the $\Lambda$DDM models, the shape of the power suppression (that is, the depth and the cut-off scale) is excellently well-captured  by our fluid prescription. Furthermore, we verified that the residuals in the structure growth parameter $S_8 \equiv \sigma_8 (\Omega_m/0.3)^{0.5}$ are always smaller than the $\sim 1.8 \%$ relative error of the $S_8$ measurement from \cite{Heymans:2020gsg}.
We thus conclude that the new WDM viscous fluid approximation is accurate enough for our analyses.

\section{\label{sec_dcdm:appendix_cs2}Semi-analytic understanding  of the WDM sound speed}
Here we obtain a formal equation that dictates the evolution of the WDM sound speed in the synchronous gauge, $c^2_{\rm s}$. 
The first natural step is to write a dynamical equation for the normalized pressure perturbation, $\Pi_\wdm = \delta P_\wdm / \bar{\rho}_\wdm$. This can be achieved by multiplying Eq.~\eqref{delta_f_0} by $4 \pi q^2 \frac{q^2}{3\mathcal{E}_\wdm} a^{-4}$, integrating over $q$ and then using Eq.~\eqref{rho_dot_wdm}. By doing so, two higher velocity-weight integrals appear, namely 
\begin{equation}
\delta \mathcal{P}_\wdm \equiv \frac{4 \pi}{3 a^4} \int_0^\infty dq  \frac{q^6}{\mathcal{E}_\wdm^3} \Delta f_{\wdm, 0},
\end{equation}
and 
\begin{equation}
(\bar{\rho}_\wdm+\bar{P}_\wdm) \Theta_\wdm \equiv \frac{4 \pi k}{a^4} \int_0^\infty dq  \frac{ q^5}{\mathcal{E}_\wdm^2} \Delta f_{\wdm, 1}.
\end{equation}
The variable $\Theta$ was already discussed in Ref.~\cite{Lesgourgues:2011rh} in the context of massive neutrinos. In the relativistic limit $\delta \mathcal{P}_\wdm $ and $\Theta_\wdm $ become equal to the standard variables $\delta P_\wdm$ and $\theta_\wdm$ (Eqs.~\eqref{pij}-\eqref{thetaj}), while in the non-relativistic limit they are suppressed by a factor $(q/\mathcal{E}_\wdm)^2$ with respect to them. This means that one can write $\delta \mathcal{P}_\wdm =   \delta P_\wdm 3 \omega_p$ and $\Theta_\wdm =  \theta_\wdm 3 \omega_{\theta} $, where $\omega_p $ and $\omega_{\theta}$ are arbitrary functions, going from $1/3$ in the relativistic limit, to $0$ in the non-relativistic case. In terms of these functions, the equation for $\Pi_\wdm$ reads 
\begin{align}
\dot{\Pi}_\wdm &= -3 \mathcal{H} \Pi_\wdm \left(\frac{2}{3}-\omega_p-\omega\right) \nonumber \\
&-\omega_\theta (1+\omega) \theta_\wdm-\frac{\dot{h}\omega}{6} \left[5-\frac{\mathfrak{p}_\wdm}{p_\wdm}\right] \nonumber \\
&+a\Gamma \frac{\bar{\rho}_\dcdm}{\bar{\rho}_\wdm} \left[\frac{\varepsilon^2}{(1-\varepsilon)}\frac{\delta_\dcdm}{3}-(1-\varepsilon)\Pi_\wdm\right]
\end{align}
\begin{figure}
\includegraphics[scale=0.22]{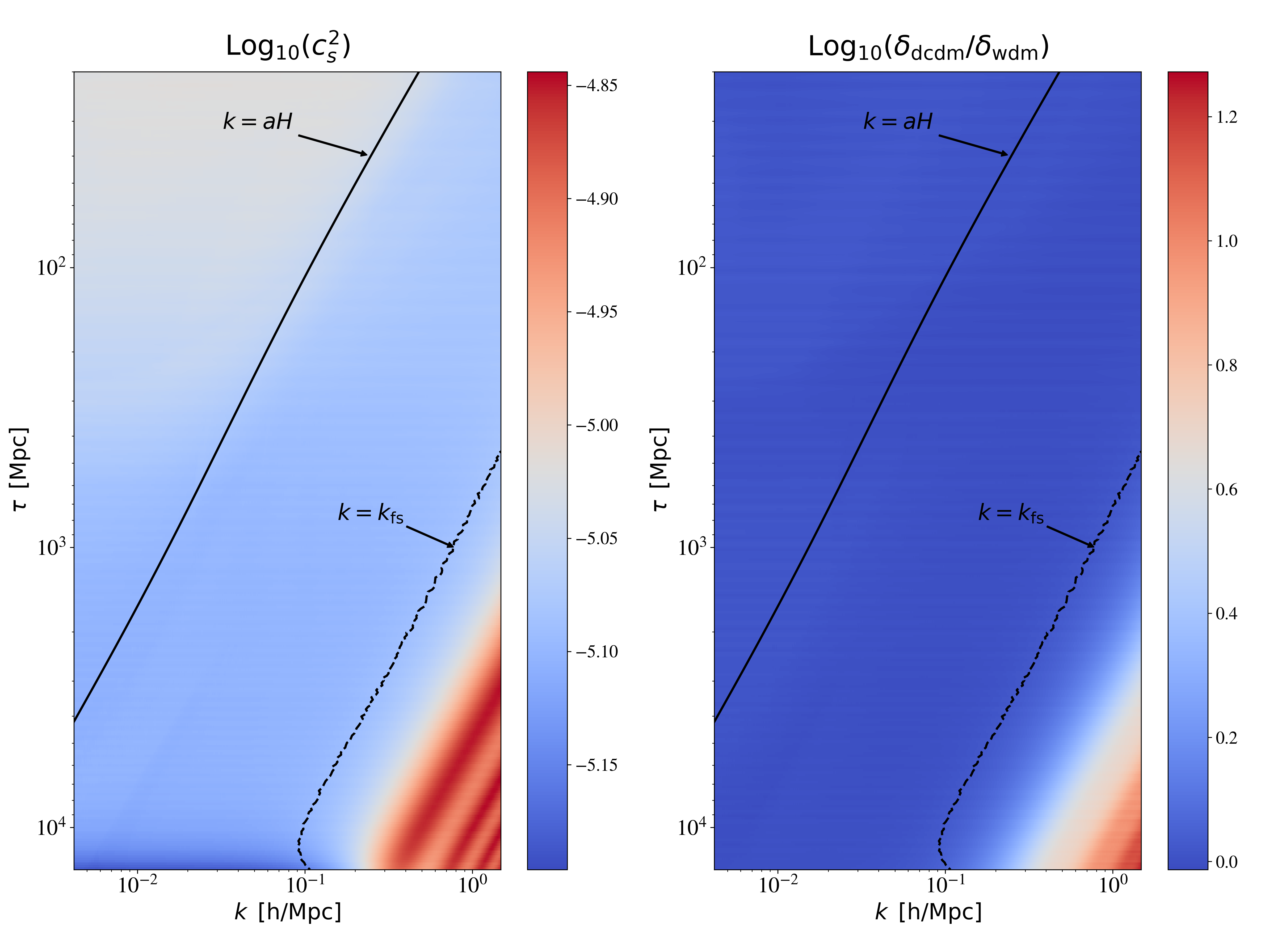}
\caption{\label{fig:cs2_and_delta_diff} {\em Left} $-$ The sound speed of the WDM species in the synchronous gauge, as a function of conformal time and wavenumber. {\em Right} $-$ The ratio between the perturbed densities of the DCDM and the WDM components, as a function of conformal time and wavenumber. The black solid and dashed lines indicate the horizon and free-streaming crossing scales, respectively.  We have set $\varepsilon = 0.007$ and $\Gamma^{-1} = 55 \ \rm{Gyrs}$.}
\end{figure}

One can convert this into an equation for the sound speed by noting that $\Pi_\wdm= c_{\rm s}^2 \delta_\wdm$, and using the continuity equation \eqref{delta_dot_wdm}. The final result reads:
\begin{align}
\frac{\partial c^2_{\rm s}}{\partial \tau}  &= -3 \mathcal{H} c^2_{\rm s} \left( \frac{2}{3}-\omega_p- c^2_{\rm s}\right) -(1+\omega) \frac{\theta_\wdm}{\delta_\wdm}(\omega_{\theta}-c^2_{\rm s})
\nonumber \\
&-\frac{\dot{h}}{2\delta_\wdm} \left[ \frac{\omega}{3}\left(5-\frac{\mathfrak{p}_\wdm}{\bar{P}_\wdm}\right)-c^2_{\rm s}(1+\omega) \right] \nonumber \\
&+a \Gamma \frac{ \bar{\rho}_\dcdm}{\bar{\rho}_\wdm} \frac{\delta_\dcdm}{\delta_\wdm} \left[ \frac{\varepsilon^2}{3(1-\varepsilon)} - (1-\varepsilon)c^2_{\rm s} \right].
\label{dot_cs2_syn}
\end{align}
We remark that the previous equation is highly non-linear in the perturbed quantities, so it can easily give rise to numerical instabilities. In addition, there is no closed expression for computing $\omega_p$ and $\omega_\theta$. If these functions were scale independent, one possible approximation would be to trade them for some background functions, such as $w$ or $c_g^2$. However, calculations using the full hierarchy show that $\omega_p$ and $\omega_\theta$ exhibit a $k$-dependence similar to that of $c_s^2$. For these reasons, we do not implement Eq.~\eqref{dot_cs2_syn} in our code. \

Nonetheless, by making some simplifying assumptions, Eq.~\eqref{dot_cs2_syn} allows to qualitatively understand why there is a particular $k$-dependence of $c^2_{\rm s}$ in the decaying scenario, that is not present in the case of massive neutrinos. Let us consider the non-relativistic limit of Eq.~\eqref{dot_cs2_syn}, since data favors in general very small DR energy fractions, $\varepsilon \ll 1$. This also implies that $w_p,c^2_{\rm s}, w \ll 1$.
Let us further assume that $\omega_\theta$ and $c^2_{\rm s}$ behave similarly, so that the difference $\omega_\theta-c^2_{\rm s}$ can be neglected. Finally, let us also restrict to sub-Hubble scales, for which the term 
$\dot{h}/\delta_\wdm$ is very small and can be also neglected. 
In this case, Eq.~\eqref{dot_cs2_syn} reduces to
\begin{align}
\frac{\partial c^2_{\rm s}}{\partial \tau}  &= -2 \mathcal{H} c^2_{\rm s} \ -a \Gamma \frac{\bar{\rho}_\dcdm}{\bar{\rho}_\wdm} \frac{\delta_\dcdm}{\delta_\wdm}c^2_{\rm s} .
\label{dot_cs2_syn2}
\end{align}
In absence of the decay term, we see that the sound speed dilutes as $c^2_{\rm s} \propto a^{-2}$, which is a well-known result for massive neutrinos. This dilution can be compensated by the presence of the decay term, leading to a $c^2_{\rm s} \sim \rm{cte}$, \textsl{as long as the ratio $\delta_\dcdm/\delta_\wdm$ doesn't change}. In practice, for scales and times such that $k < k_{\rm fs} (\tau)$, we have $\delta_\dcdm/\delta_\wdm=1$. In this regime, the sound speed $c^2_{\rm s}$ is well approximated by the adiabatic sound speed $c_g^2$.
However, when $k > k_{\rm fs} (\tau)$, $\delta_\wdm$ oscillates and starts to become suppressed with respect to $\delta_\dcdm$, which leads to oscillatory features and a small enhancement in the evolution of $c^2_{\rm s}$. This is visible in Fig. \ref{fig:cs2_and_delta_diff}, where we have plotted $c^2_{\rm s}$ and $\delta_\dcdm/\delta_\wdm$ in the $k-\tau$ plane using the full Boltzmann hierarchy, for the best-fit parameters from the combined analysis of Ref.~\cite{Abellan:2020pmw} (i.e., setting $\varepsilon = 0.007$ and $\Gamma^{-1} = 55 \ \rm{Gyrs}$).

~\\
One can see that this $k$-dependent effect appears only because of the coupling term in Eq.~\eqref{dot_cs2_syn2}, which is not present for massive neutrinos. This also justifies why the sound speed $c^2_{\rm s}$ is well approximated by a background function such as $c_g^2$ in the case of massive neutrinos. In the case of the WDM daughter species, the approximation   $c^2_{\rm s} \simeq c_g^2 $ will only work when $k < k_{\rm fs} (\tau)$. This motivates the use of the fitting formula introduced in Eq.~\eqref{sync_sound_speed}, that accounts for the small enhancement at scales smaller than the free-streaming scale. While this simple fitting formula is not able to capture the oscillatory features described previously, it leads to results that are accurate enough for all the observables analysed in this work.

\section{\label{sec_dcdm:appendix_planck}Consistency between the use of the {\emph{Planck}} TTTEEE `lite' and `full' likelihoods}

\begin{figure}
    \centering
    \includegraphics[scale=0.42]{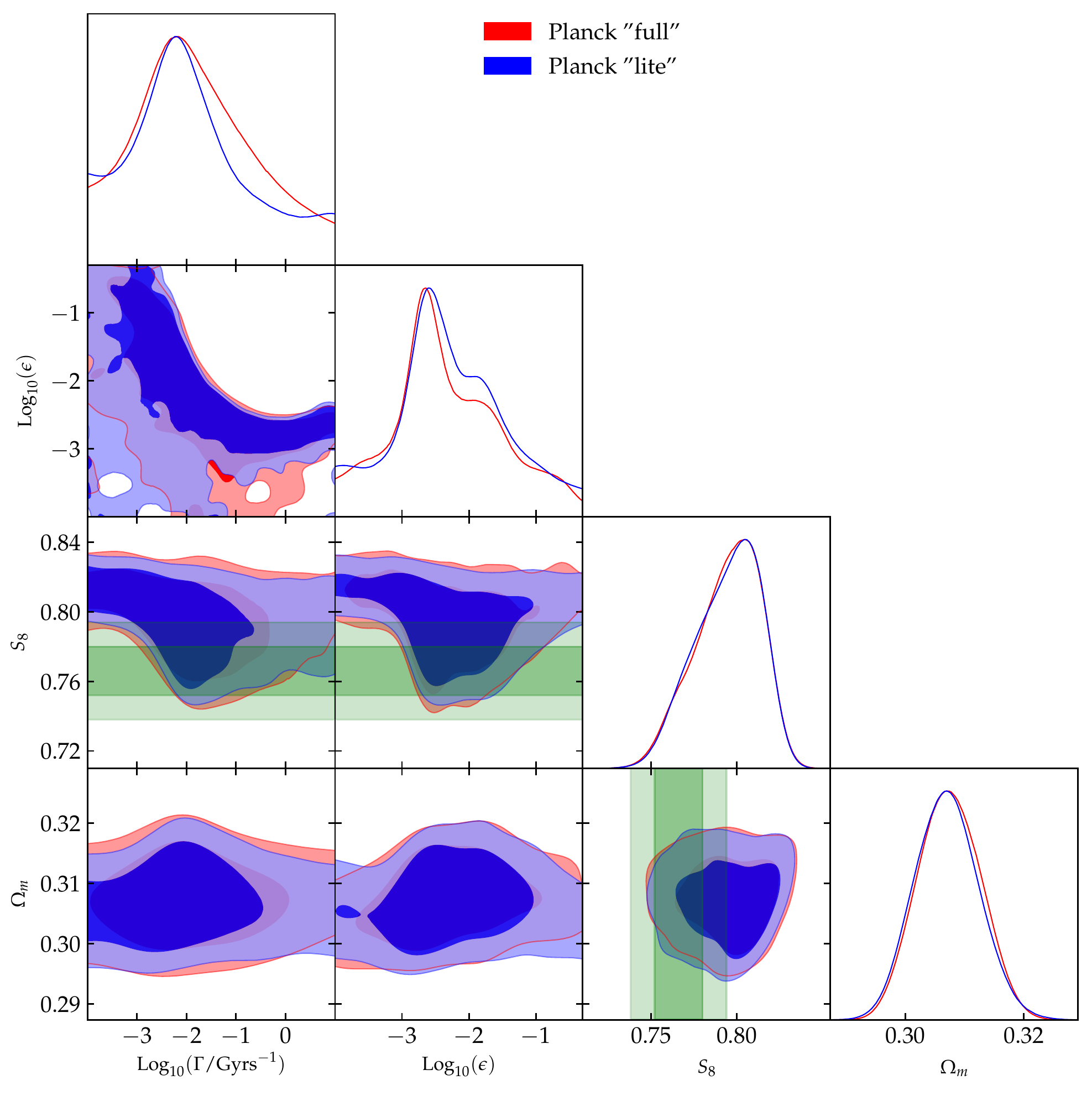}
    \caption{2D posterior distribution for a subset of parameters obtained by using the {\emph{Planck}} `full' likelihood, compared to those from the {\emph{Planck}} `lite' analysis.}
    \label{fig:plk_full}
\end{figure}

\begin{table}
 \begin{tabular}{|l|c|c|} 
 \hline
 Parameter & $Planck$ `full' & $Planck$ `lite' \\ 
 \hline
$100 \ \omega_b $								&  $2.243_{-0.015}^{+0.012}$	 & $2.246 \pm 0.013$            \\
$\Omega_{\rm dcdm}^{\rm ini}$ 					&  $0.2585_{-0.0048}^{+0.0054}$  & $0.2581_{-0.0054}^{+0.005}$ \\
$ H_0 /[{\rm km/s/Mpc}]$ 						&  $67.88_{-0.44}^{+0.36}$  & $67.92_{-0.42}^{+0.43}$      \\
$\text{ln}(10^{10} A_s)$						&  $3.046_{-0.015}^{+0.015}$  & $3.048_{-0.016}^{+0.014}$    \\
$n_s$											&  $0.9667_{-0.0039}^{+0.0036}$  & $0.9682 \pm 0.0037 $        \\
$\tau_{\rm reio}$								&  $0.0562_{-0.0072}^{+0.0073}$  & $0.0570_{-0.0077}^{+0.0071}$\\
$\text{log}_{10} (\Gamma/[{\rm Gyr}^{-1}])$		&  $-1.84_{-1.4}^{+1}$  & $-1.89_{-1.5}^{+0.82}$       \\
$\text{log}_{10} (\varepsilon)$					&  $-2.26_{-0.82}^{+0.84}$  & $-2.28_{-0.78}^{+0.8}$       \\
$\Omega_{\rm m}$								&  $0.3075_{-0.0049}^{+0.0057}$  & $0.3071_{-0.0057}^{+0.0053}$\\
$S_8$										    &  $0.794_{-0.018}^{+0.024}$  & $0.795_{-0.016}^{+0.024}$    \\
\hline
\end{tabular}
\caption{The mean $\pm 1\sigma$ errors of the cosmological parameters from our BAO + SNIa + \emph{Planck} + $S_8$ analysis, obtained by using both the \emph{Planck} `full' and `lite' likelihoods.}
\label{tab:plk_full}
\end{table}

Given that for all our analyses we made use of the `lite' version of the {\emph{Planck}} likelihood, in Fig.~\ref{fig:plk_full} and Tab.~\ref{tab:plk_full}
we show that our conclusions are not affected by such a choice, by comparing the results from our main run with the predictions from a `full' {\emph{Planck}} analysis.

\section{\label{sec_dcdm:chi2}Best-fit $\chi^2$ per experiment}
In Tab.~\ref{tab:chi2_all} we report all $\chi^2_{\rm min}$'s obtained with the {\sc Minuit} algorithm \cite{James:1975dr} through the {\sc iMinuit} python package for the various model and data-set combinations considered in this work.

\begin{table*}
\scalebox{1}{
\begin{tabular}{|l|c|c|c|c|c|c|c|c|c|c|c|}
\hline
& BAO+SNIa & +{\emph{Planck}} & \multicolumn{5}{c|}{+{\emph{Planck}} w/ $S_8$} & \multicolumn{2}{c|}{+SPTpol} & \multicolumn{2}{c|}{+ACTPol} \\ 
& & w/o $S_8$ & \multicolumn{3}{c|}{} & w/ $A_{\rm lens}$ & $\varepsilon=0.05$ & w/o $S_8$  & w/ $S_8$& w/o $S_8$ & w/ $S_8$ \\
\hline
Pantheon SNIa & 1026.9 & 1028.2 & 1027.5 & 1026.8 &  1029.2  & 1028.0 & 1027.1 & 1027.0 & 1026.9 & 1027.0 & 1026.84\\
BAO+FS~BOSS DR12 & -- & 6.63 & 7.06 & 6.93 & 7.11  & 6.14 & 6.59 & 5.94 & 6.17 & 6.78 &6.66 \\
BAO~BOSS DR12 & 3.52 & -- & -- & -- & -- & -- & -- & -- & -- & -- & --\\
BAO~BOSS low$-z$ & 1.84 & 1.20 & 1.22 & 1.18 & 1.24 & 1.83 & 1.28 & 1.58 & 1.79 & 1.17 & 1.34\\
BAO~eBOSS DR14 & 4.29 & 4.94 & 4.91 & 4.94 & 4.83 & 4.52 & 4.88 & 4.68 & 4.53 & 4.97 & 4.77\\
{\emph{Planck}}~high$-\ell$ TT,TE,EE `lite' & -- & 584.8 & 585.9 & 585.3 & 586.9 & 577.7 & 587.5 & -- & -- & 589.216 & 590.545  \\
{\emph{Planck}}~high$-\ell$ TT `lite' & -- & -- & -- & -- & -- & -- & -- & 207.8 & 207.8 & -- & -- \\
{\emph{Planck}}~low$-\ell$ EE & -- & 396.9 & 396.9 & 397.2 & 396.3 & 395.7 & 396.3 & 395.8 & 396.1 & 396.2 & 397.15\\
{\emph{Planck}}~low$-\ell$ TT & -- & 23.1 & 23.1 & 23.2 & 23.0 & 22.1 & 23.0 & 22.3 & 22.1 & 22.6 &  22.46\\
{\emph{Planck}}~lensing & -- & 8.78 & 9.12 & 8.88 & 9.47 & 8.53 & 9.83  & -- & -- & 8.8 &8.94 \\ 
SPTpol~high$-\ell$ TE,EE & -- & -- & -- & -- & -- & -- & -- & 145.9 & 145.6 & -- & -- \\
SPTpol~lensing & -- & -- & -- & -- & -- & -- & -- & 5.43 & 5.93& -- & -- \\
ACTPol & -- & -- &-- &-- &-- &-- &-- &-- &-- &238.235 &237.359 \\
KiDS+BOSS+2dFLens & -- & -- & 0.0003 & -- & -- & 0.0097 & 0.98 & -- & 0.0015 & -- & \\
DES & -- & -- & -- & 0.19 & -- & -- & -- & -- & -- & -- & --  \\
KiDS+Viking+DES & -- & -- & -- & -- & 0.20 & -- & -- & -- & -- & -- & -- \\
\hline
total $\chi^2$ & 1036.6 & 2053.4 & 2055.0 & 2054.8 & 2055.9 & 2043.2 & 2057.6 & 1816.3 & 1816.8 & 2294.8 &  2296.2 \\
\hline
\end{tabular}}
\caption{Best-fit $\chi^2$ per experiment (and total) for all the $\Lambda$DDM analyses performed in this work.}
\label{tab:chi2_all}
\end{table*}

\section{\label{sec_dcdm:appendix_comparison} Comparison with the Planck constraints from Ref. \citep{Clark:2020miy} }
\begin{figure}[t]
    \centering
    \includegraphics[scale=0.4]{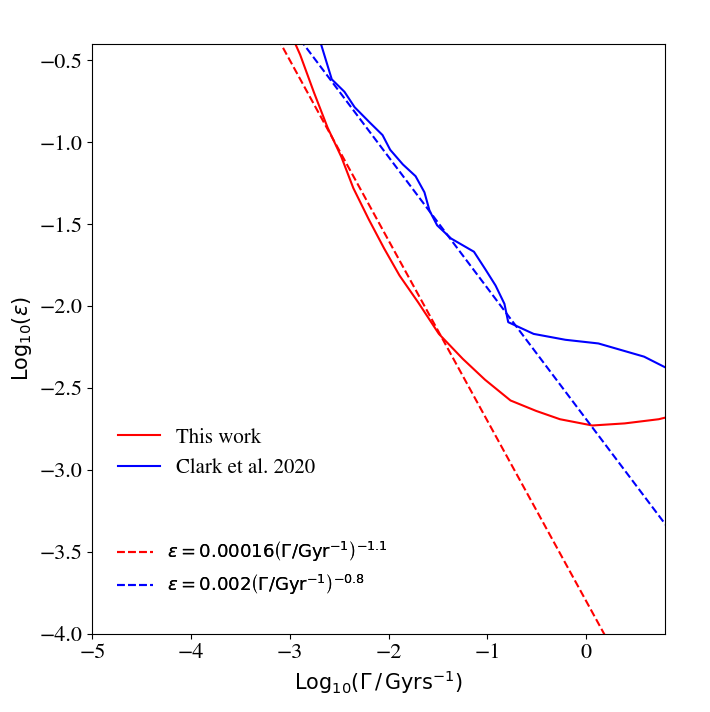}
    \caption{Comparison between the $2\sigma$ exclusion bounds (solid lines) from the \emph{Planck} analysis of Ref. \citep{Clark:2020miy} and our \emph{Planck}+BAO+SNIa analysis. In each case, the dashed line indicates a fit that roughly describes the $2\sigma$ limit in the range $\Gamma \sim 10^{-3}-10^{-1} \ \text{Gyrs}^{-1}$. }
    \label{fig:constraints_comparison}
\end{figure}

Here we carry out an explicit comparison of our constraints with those of Ref. \citep{Clark:2020miy}, which performed an analysis of the $\Lambda$DDM model against {\em Planck} data, neglecting the perturbations of the warm daughter particles. As shown in Fig. \ref{fig:constraints_comparison}, we find that the constraints on the $\Lambda$DDM models are up to (roughly) one order of magnitude stronger when our improved treatment is considered.

\section{Results with a linear prior on $\Gamma$ and $\varepsilon$}

\begin{figure}[h]
    \centering
    \includegraphics[scale=0.4]{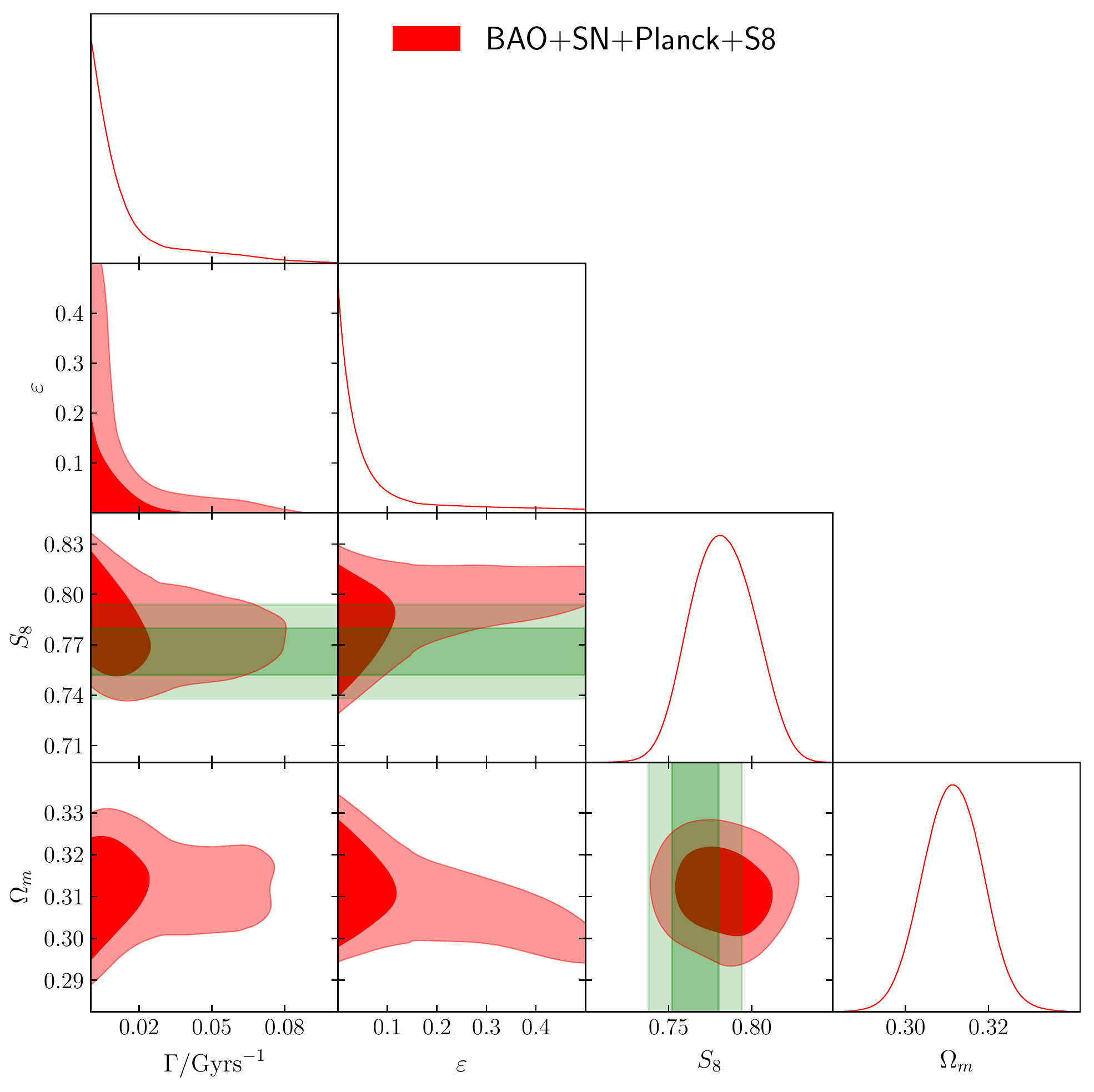}
    \includegraphics[scale=0.4]{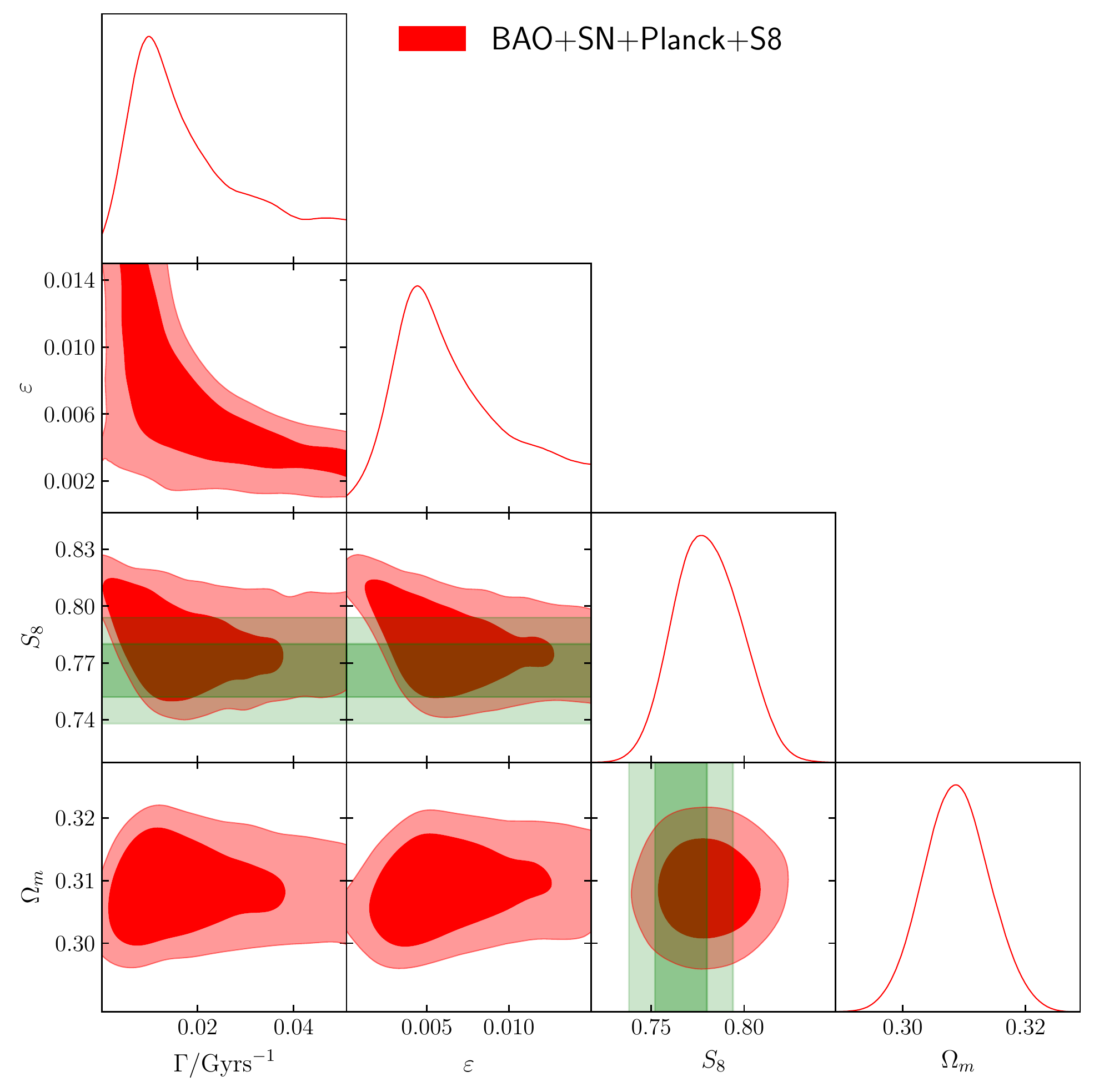}

    \caption{Reconstructed 2D posteriors of a
    BAO + SNIa + \emph{Planck} + $S_8$ (from KiDS+BOSS+2dFLens) analysis,
    with linear priors and sampling either with the original prior range (upper panel) or within a restricted prior range (lower panel).}
    \label{fig:linpriors}
\end{figure}

{
In our baseline analysis we have made use of log-prior on $\varepsilon$ and $\Gamma$, to ease comparison with earlier works \cite{vattis_late_2019,Clark:2020miy} who adopted the same choice. Here we present results using linear priors on the DDM parameters.
Let us however stress that the use of a linear prior is less informative than adopting a logarithmic one. That is because a linear prior carries a scale (due to the large error bars used in the proposal distribution of the MCMC sampler), so that it is hard for the sampler to detect fine structure over 4 orders of magnitude by using a linear scale, in particular at very small values. In other words, given the difference between the scale of the upper limit on $\Gamma /\rm{Gyrs}^{-1}$ an $\varepsilon$ ($\sim10^{-1}$) and that of the lower limit ($\sim10^{-3}$)), it is very difficult to accurately reconstruct the parameter space with a linear prior. Such a difficulty is illustrated in Fig.~\ref{fig:linpriors}, where we provide the results of two linear-prior analyses: the upper panel corresponding to the original prior range, the lower panel corresponding to a more restricted range: $\varepsilon \in [0.0001,0.015]$ and $\Gamma/\rm{Gyrs}^{-1} \in [0.0001,0.05]$. \looseness=-1 While in the latter case $\varepsilon$ and $\Gamma$ are detected at the 2$\sigma$ level -- similarly to the log-prior results -- the former case weighs in favor of larger $\varepsilon$ values, so that one would uncorrectly deduce an upper limit only.
}

\newpage
\bibliography{dcdm}

\end{document}